\documentclass[twocolumn,showpacs,amssymb,prd,superscriptaddress,nofootinbib]{revtex4-1}

\usepackage{graphicx,epsf, epsfig, amssymb}
\usepackage{bm}
\usepackage{longtable}
\usepackage{color}
\usepackage{hyperref}
\usepackage{amsfonts,amsmath,amssymb,mathrsfs}

\def\be{\begin{equation}}
\def\ee{\end{equation}}
\def\ba{\begin{eqnarray}}
\def\ea{\end{eqnarray}}
\newcommand{\nn}{\nonumber}

\newcommand{\ISCO}{{\mbox{\tiny ISCO}}}

\newcommand{\bgaln}{\begin{align}}
\newcommand{\bgeq}{\begin{equation}}

\newcommand{\Rmnum}[1]{\expandafter\@slowromancap\romannumeral #1@}

\begin{document}

\title {Accuracy of the post-Newtonian approximation. II. Optimal asymptotic
  expansion of the energy flux for quasicircular, extreme mass-ratio inspirals
  into a Kerr black hole.}

\author{Zhongyang Zhang}
\affiliation{Department of Physics and Astronomy, The University of
  Mississippi, University, MS 38677, USA}

\author{Nicol\'as Yunes}
\affiliation{Department of Physics and MIT Kavli Institute, 77 Massachusetts
  Avenue, Cambridge, MA 02139, USA}

\author{Emanuele Berti}
\affiliation{Department of Physics and Astronomy, The University of
  Mississippi, University, MS 38677, USA}
\affiliation{California Institute of Technology, Pasadena, CA 91109, USA}

\date{\today}

\begin{abstract}
We study the effect of black hole spin on the accuracy of the post-Newtonian
approximation. We focus on the gravitational energy flux for the
quasicircular, equatorial, extreme mass-ratio inspiral of a compact object
into a Kerr black hole of mass $M$ and spin $J$. For a given dimensionless
spin $a\equiv J/M^2$ (in geometrical units $G=c=1$), the energy flux depends
only on the orbital velocity $v$ or (equivalently) on the Boyer-Lindquist
orbital radius $r$. We investigate the formal region of validity of the Taylor
post-Newtonian expansion of the energy flux (which is known up to order $v^8$
beyond the quadrupole formula), generalizing previous work by two of us. The
{\emph{error function}} used to determine the region of validity of the
post-Newtonian expansion can have two qualitatively different kinds of
behavior, and we deal with these two cases separately. We find that, at any
fixed post-Newtonian order, the edge of the region of validity (as measured by
$v/v_{\ISCO}$, where $v_{\ISCO}$ is the orbital velocity at the innermost
stable circular orbit) is only weakly dependent on $a$. Unlike in the
nonspinning case, the lack of sufficiently high order terms does not allow us
to determine if there is a convergent to divergent transition at order
$v^{6}$. Independently of $a$, the inclusion of angular multipoles up to and
including $\ell=5$ in the numerical flux is necessary to achieve the level of
accuracy of the best-known ($N=8$) PN expansion of the energy flux.
\end{abstract}
\maketitle

\section{Introduction}

Binaries of compact objects, such as black holes (BHs) and/or neutron stars,
are one of the main targets for gravitational-wave (GW) observations. When the
binary members are widely separated, their slow inspiral can be well-described
by the post-Newtonian (PN) approximation, a perturbative {\emph{asymptotic}}
expansion of the ``true'' solution of the Einstein equations. The small
expansion parameter in the PN approximation is $v/c$, where $v$ is the orbital
velocity of the binary and $c$ is the speed of light. Asymptotic expansions,
however, must be used with care, as the inclusion of higher-order terms does
not necessarily lead to an increase in accuracy.  Therefore one would like
to determine the optimal order of expansion and the formal region of validity
of the PN asymptotic series~\cite{Bender:1978,Yunes:2008tw}, i.e.~the order
and region inside which the addition of higher order terms increases the
accuracy of the approximation in a convergent fashion.

In a previous paper (\cite{Yunes:2008tw}, henceforth Paper I), two of us
investigated the accuracy of the PN approximation for quasicircular,
nonspinning (Schwarzschild), extreme mass-ratio inspirals (EMRIs). By
comparing the PN expansion of the energy flux to numerical calculations in the
perturbative Teukolsky formalism, we concluded that (i) the region of validity
of the PN expansion is largest at relative 3PN order -- i.e., order $(v/c)^6$
(throughout this paper, a term of ${\cal{O}}(v^{2N})$ is said to be of $N$PN
order); and (ii) the inclusion of higher multipoles in numerical calculations
is necessary to improve the agreement with PN expansions at large orbital
velocities.  The fact that the region of validity is largest at 3PN could be a
hint that the series actually {\em diverges} beyond 3PN order, at least in the
extreme mass-ratio limit.

This paper extends our study to EMRIs for which the more massive component is
a {\emph{rotating}} (Kerr) BH. The present analysis focuses on the effect of
the BH spin on the accuracy of the PN expansion. We generalize the methods
presented in Paper I to take into account certain pathological behaviors of
the {\emph{error function}}, used to determine the region of validity. This
generalization may also be applicable to comparable-mass systems.

A surprising result we find is that the edge of the region of validity (the
maximum velocity beyond which higher-order terms in the series cannot be
neglected), normalized to the velocity at the innermost stable circular orbit,
is weakly dependent on the Kerr spin parameter $a$. In fact, this edge is
roughly in the range $v/v_{\ISCO} \in [0.3,0.6]$ for almost all PN orders,
irrespective of $a$. This suggests, perhaps, that the ratio $v/v_{\ISCO}$ is a
better PN expansion parameter than $v/c$, when considering spinning BHs.

Another surprising result is related to the behavior of the edge of the region
of validity as a function of PN order. In the nonspinning case, two of us
found that beyond 3PN order, ${\cal{O}}(v^{6}/c^{6})$, this edge seemed to
consistently decrease with PN order~\cite{Yunes:2008tw}.  This was studied up
to ${\cal{O}}(v^{11}/c^{11})$, the largest PN order known for the nonspinning
case. In the spinning case, however, the series is known only up to
${\cal{O}}(v^{8}/c^{8})$, and we are thus unable to conclusively determine if
the trend found in the nonspinning case persists. Higher-order calculations
will be necessary to draw more definite conclusions.

Numerical (or in this case, perturbative) calculations of the energy flux rely
on multipolar decompositions of the angular dependence of the radiation. By
comparing the convergence of the multipolar decomposition to the convergence
of the PN expansion of the energy flux, we find that for $v/c\sim 0.1$ the
inclusion of multipoles up to and including $\ell=5$ seems necessary to
achieve the level of accuracy of the best-known ($N=8$) PN expansion of the
flux. These conclusions are also largely independent of the spin parameter
$a$.

The rest of this paper is organized as follows. In Section \ref{sec:flux} we
present the energy flux radiated by quasicircular, equatorial Kerr EMRIs in
the adiabatic approximation, as computed in PN theory
\cite{Poisson:1995vs,Tagoshi:1996gh,Tanaka:1997dj} and with accurate
frequency-domain codes in BH perturbation theory
\cite{2009GWN.....2....3Y,Yunes:2009ef,Yunes:2010zj}. In Section
\ref{sec:validity} we discuss the region of validity of the PN approximation
in terms of the normalized orbital velocity $v/v_{\ISCO}$ and of the
normalized orbital radius $r/r_{\ISCO}$, where ISCO stands for the innermost
stable circular orbit. We consider both corotating and counterrotating
orbits. In Section \ref{sec:multipoles} we study the number of multipolar
components that must be included in the {\em numerical} flux in order to
achieve sufficient accuracy. Finally, in Section \ref{sec:concl} we present
our conclusions. We follow the same notation as in Paper I. In particular,
from now on we will use geometrical units ($G=c=1$).

\section{\label{sec:flux}Energy flux for quasicircular, equatorial EMRIs in
  Kerr: numerical and PN results}

In the PN approximation, the GW energy flux radiated to
infinity by a test particle in a circular orbit and on the equatorial plane of a
Kerr BH is given by~\cite{Tagoshi:1996gh,Tanaka:1997dj,Poisson:1995vs}
\ba\label{PNexp}
F^{(N)}=F_{\rm Newt}\left[\sum_{k=0}^N(a_k+b_k\,\ln v)\,v^k\right]\,.
\ea
This flux is known up to $N=8$ when including spins, and up to $N=11$ in the
nonspinning case. The leading (Newtonian) contribution\footnote{Notice that
  there is a typo in Eq.~(18) of Paper I.} is
\be
F_{\rm Newt}=\frac{32}{5}\frac{\mu^2}{M^2}v^{10}\,,
\ee
where $\mu$ and $M$ are the test particle mass and Kerr BH mass,
respectively.  As we are here interested in the accuracy of the PN
approximation, we will ignore the flux of energy going into the horizon, which
cannot always be neglected when building waveform templates.

The expansion coefficients $a_k$ and $b_k$ contain both spin-independent and
spin-dependent terms, where the dimensionless spin parameter $a$ is related
to the Kerr BH spin angular momentum via $J = a M^{2}$. These coefficients
can be found in Eq.~(G19) of \cite{Tagoshi:1996gh}, so we do not list them
explicitly\footnote{See also their Eq.~(3.40), that provides a similar
  expansion in terms of the PN orbital velocity parameter
  $v'=\sqrt{M/r_0}$.}. Note that logarithmic terms only appear at 3PN and 4PN
(i.e., $b_6\neq 0$ and $b_{8}\neq 0$), and that the $(a_{k},b_{k})$ for
$8<k\leq 11$ are known {\it only} in the spin-independent limit.

Throughout this paper $v$ is the orbital velocity, defined as $v\equiv
(M\Omega)^{1/3}$ (where $\Omega$ is the small body's orbital frequency), and
related to the Boyer-Lindquist radius $r$ by
\ba
\label{r-solve}
\frac{r}{M}=\frac{(1-av^3)^{2/3}}{v^2}\,,
\ea
whose inverse is
\be \label{kepler}
v =  \left[(r/M)^{3/2} + a\right]^{-1/3}\,.
\ee
At the ISCO we have \cite{Bardeen:1972fi}
\ba
\label{risco}
\frac{r_{\ISCO}}{M}&=&3+Z_2 - \sqrt{(3-Z_1)(3+Z_1+2Z_2)}\,,\\
Z_1&\equiv&1+(1-a^2)^{1/3}\left[(1+a)^{1/3}+(1-a)^{1/3}\right]\,,\nn\\
Z_2&\equiv&(3a^2+Z_1^2)^{1/2}\,,
\label{z1z2}
\ea
where $a>0$ ($a<0$) corresponds to corotating (counterrotating) orbits.

\begin{figure}
\centerline{\includegraphics[width=8.5cm,clip=true]{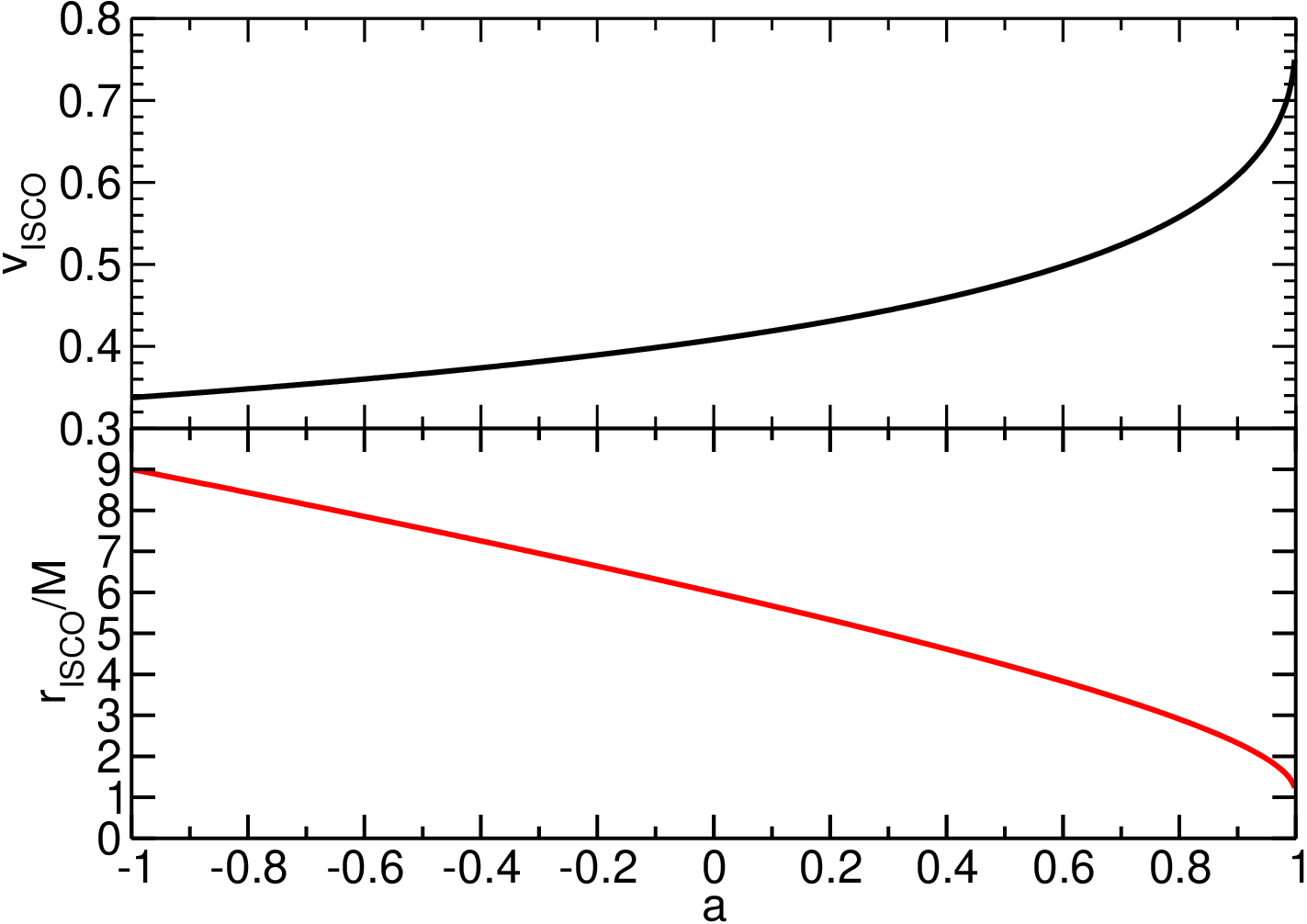}}
\caption{\label{visco} ISCO velocity ($v_{\ISCO}$, top panel) and radius
  ($r_{\ISCO}/M$, bottom panel) as a function of $a$. Here and elsewhere we
  use the convention that a negative spin parameter refers to counterrotating
  orbits.}
\end{figure}

Using Eqs.~\eqref{r-solve} and~(\ref{risco}), we also have
\ba
\frac{r}{r_{\ISCO}}=
\frac{(1-av^3)^{2/3}}{v^2\left[3+Z_2\mp\sqrt{(3-Z_1)(3+Z_1+2Z_2)}\right]}\,.
\notag
\ea
The ISCO velocity can be found by replacing $r_{\ISCO}$ in Eq.~\eqref{risco}
into the velocity-radius relation (\ref{kepler}). The velocity $v_{\ISCO}$ and
the radius $r_{\ISCO}/M$ are displayed graphically in
Fig.~\ref{visco}. Observe that, although $r_{\ISCO} \to M$ as $a \to 1$,
$v_{\ISCO}$ is bounded by $2^{-1/3}\simeq 0.79$.

\begin{figure*}[htb]
\begin{tabular}{cc}
\includegraphics[width=8.5cm,clip]{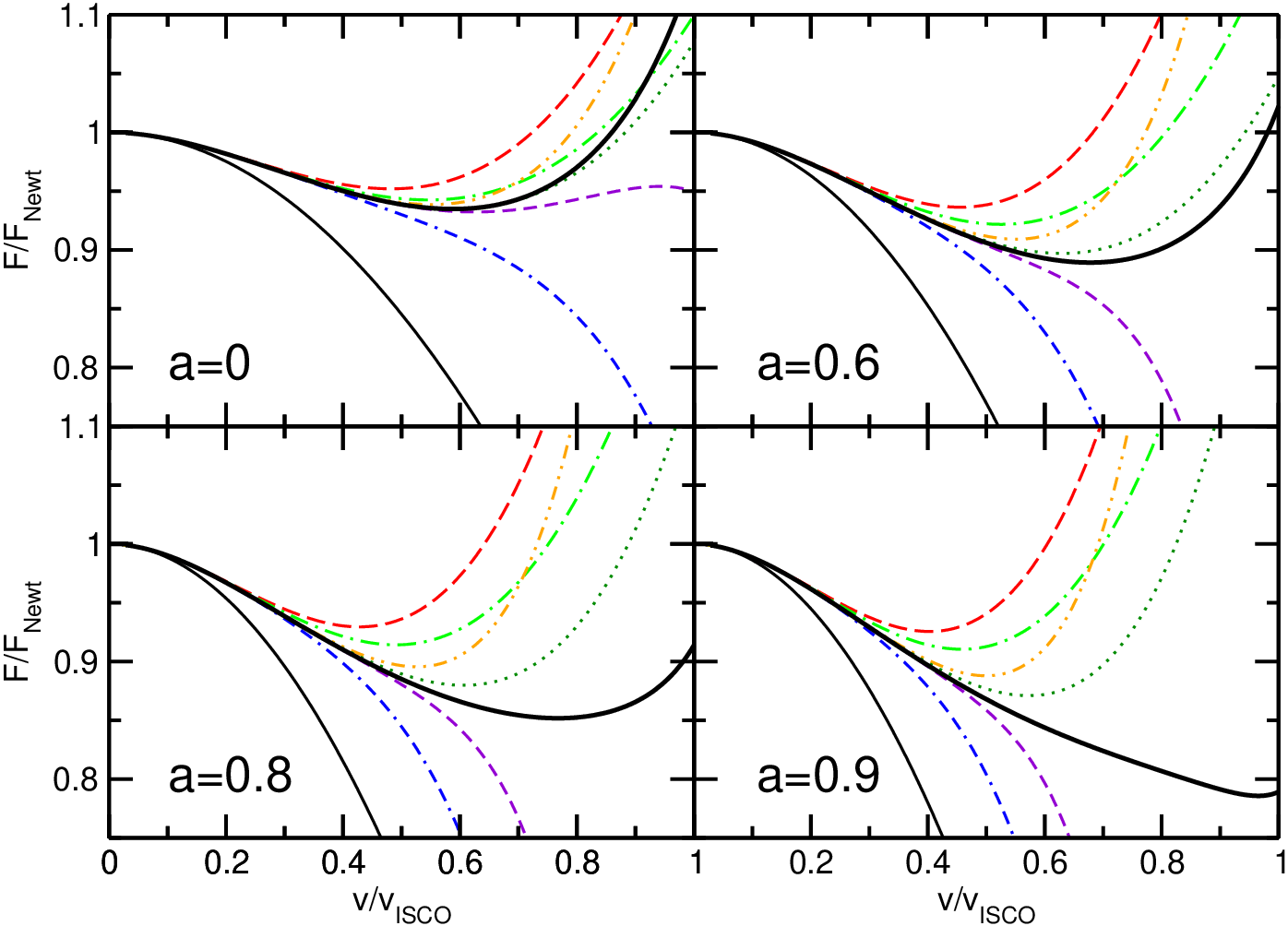}&
\includegraphics[width=8.5cm,clip]{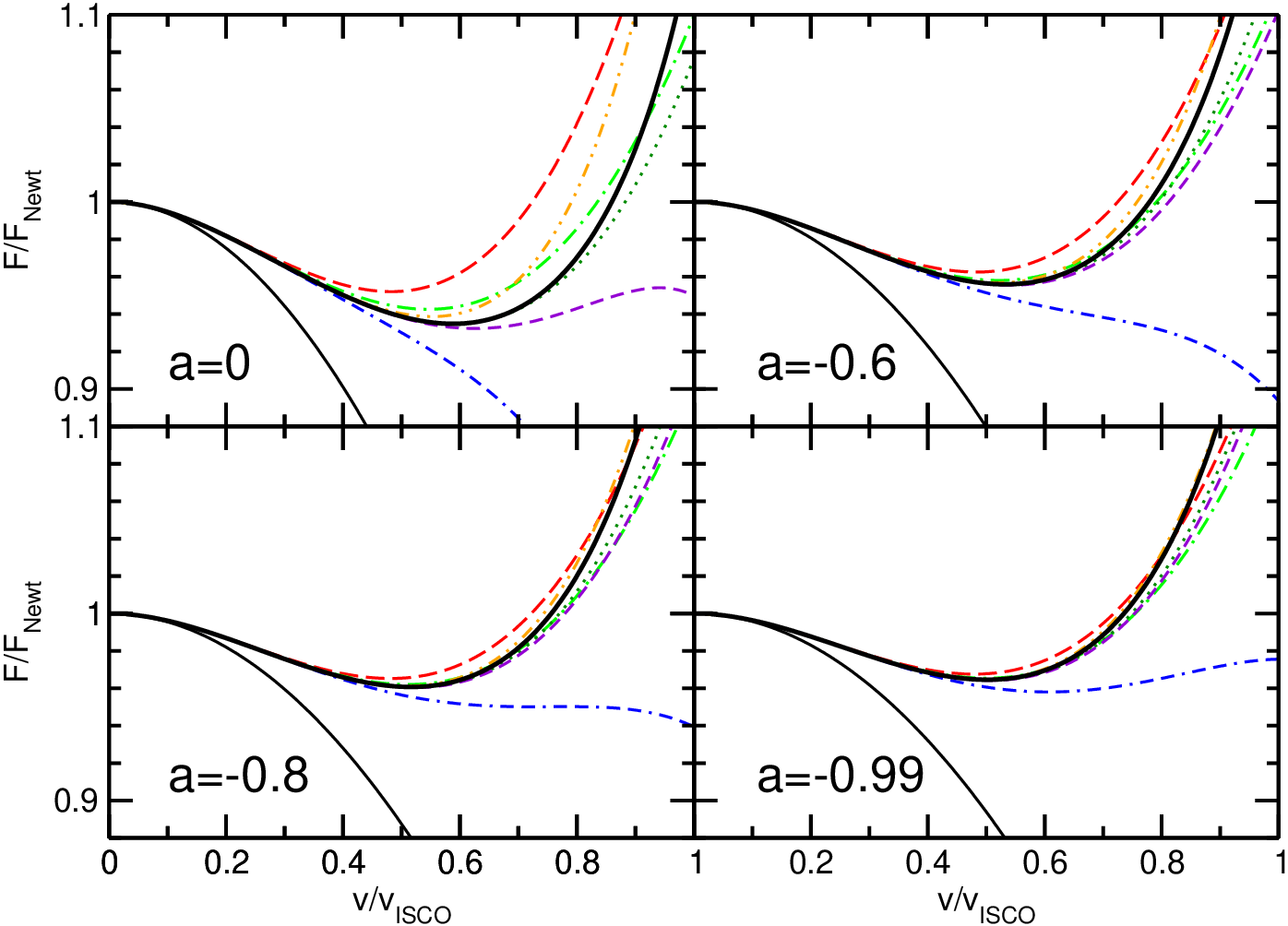}\\
\end{tabular}
\caption{\label{result} Gravitational energy flux (normalized to $F_{\rm
    Newt}$) as a function of the normalized orbital velocity, $v/v_{\ISCO}$.
    The left panel is for corotating orbits, and the right panel for
  counterrotating orbits. Different insets refer to different spin parameters
  $a$, as indicated. The thick black line is the numerical flux. Other
  linestyles refer to different PN approximations: $F^{(2)}$ (thin black),
  $F^{(3)}$ (long-dashed red), $F^{(4)}$ (dash-dotted green), $F^{(5)}$
  (dash-dash-dotted blue), $F^{(6)}$ (dash-dot-dotted orange), $F^{(7)}$
  (dotted dark green), $F^{(8)}$ (short-dashed violet).}
\end{figure*}

The rigorous definition of velocity is a tricky business in general
relativity. We have here chosen to define velocity in a quasi-Newtonian
fashion, in terms of the angular velocity and Kepler's law. One can think of
this velocity as that which would be measured by an observer at spatial
infinity. On the other hand, one can also study the velocity measured by an
observer in the neighborhood of the BH and that is rotating with the geometry;
this quantity would differ from $v$ in Eq.~\eqref{kepler}, and, in fact, its
associated $v_{\ISCO}$ would tend to $1/2$ in the limit $a \to 1$ (see
eg. Eq. $(3.11b)$ in~\cite{Bardeen:1972fi}). This shows that the $a \to 1$
limit is very delicate, and the precise value of the velocity field is an
observer-dependent (and a non-invariant) quantity. However, once a definition
is chosen, the velocity is a perfectly good quantity to parametrize the
structure of the PN series.

A first guess at the asymptotic behavior of this series can be obtained by
simply plotting different PN approximants $F^{(N)}$ and comparing them with
high-accuracy, numerical results for the energy flux, obtained from a
frequency-domain Teukolsky code (see \cite{Poisson:1995vs,Mino:1997bx} for
early work in the Schwarzschild case, and Fig.~9 in \cite{Pan:2010hz} for a
related discussion in the Kerr context).  The numerical results used in this
comparison are the same as those used in
Refs.~\cite{2009GWN.....2....3Y,Yunes:2009ef,Yunes:2010zj} to study the
accuracy of a resummed effective-one-body version of the PN approximation to
model EMRIs. They consist of numerical fluxes, evaluated for spin parameters
ranging from $a=0$ to $a=0.9$ in steps of $\Delta a=0.1$ (in fact, we also
have access to the counterrotating flux for $a=-0.99$). The typical accuracy
of these fluxes is better than one part in $10^{10}$ for all velocities and
spins. We refer the reader to Section IIB of \cite{Yunes:2010zj} for a more
detailed description of the code.

Figure~\ref{result} compares the different PN approximations to the numerical
flux. The left panel refers to corotating orbits, and the right panel to
counterrotating orbits. Different insets correspond to different values of the
BH spin, and different linestyles represent different orders in the PN
expansion. As stated earlier, in this figure and in the rest of this paper, we
neglect energy absorption by the BH.  Observe that, as first noted by Poisson
in the Schwarzschild case \cite{Poisson:1995vs}, the behavior of the PN
expansion is quite erratic. For any given $a$, rather than converging
monotonically, higher-order approximations keep undershooting and overshooting
with respect to the ``exact'' numerical result. This oscillatory behavior is
quite typical of asymptotic expansions, and it has been studied in depth,
especially for extreme mass-ratio inspirals into nonrotating BHs
\cite{Poisson:1995vs,Mino:1997bx}. Various authors proposed different schemes
to accelerate the convergence of the PN expansion, including Pad\'e
resummations \cite{Damour:1997ub,Damour:2000zb,Mroue:2008fu} and the use of
Chebyshev polynomials \cite{Porter:2007vk}. The asymptotic properties of these
resummation techniques are an interesting topic for future study.

\begin{figure*}[htb]
\begin{tabular}{cc}
\includegraphics[width=8.5cm,clip]{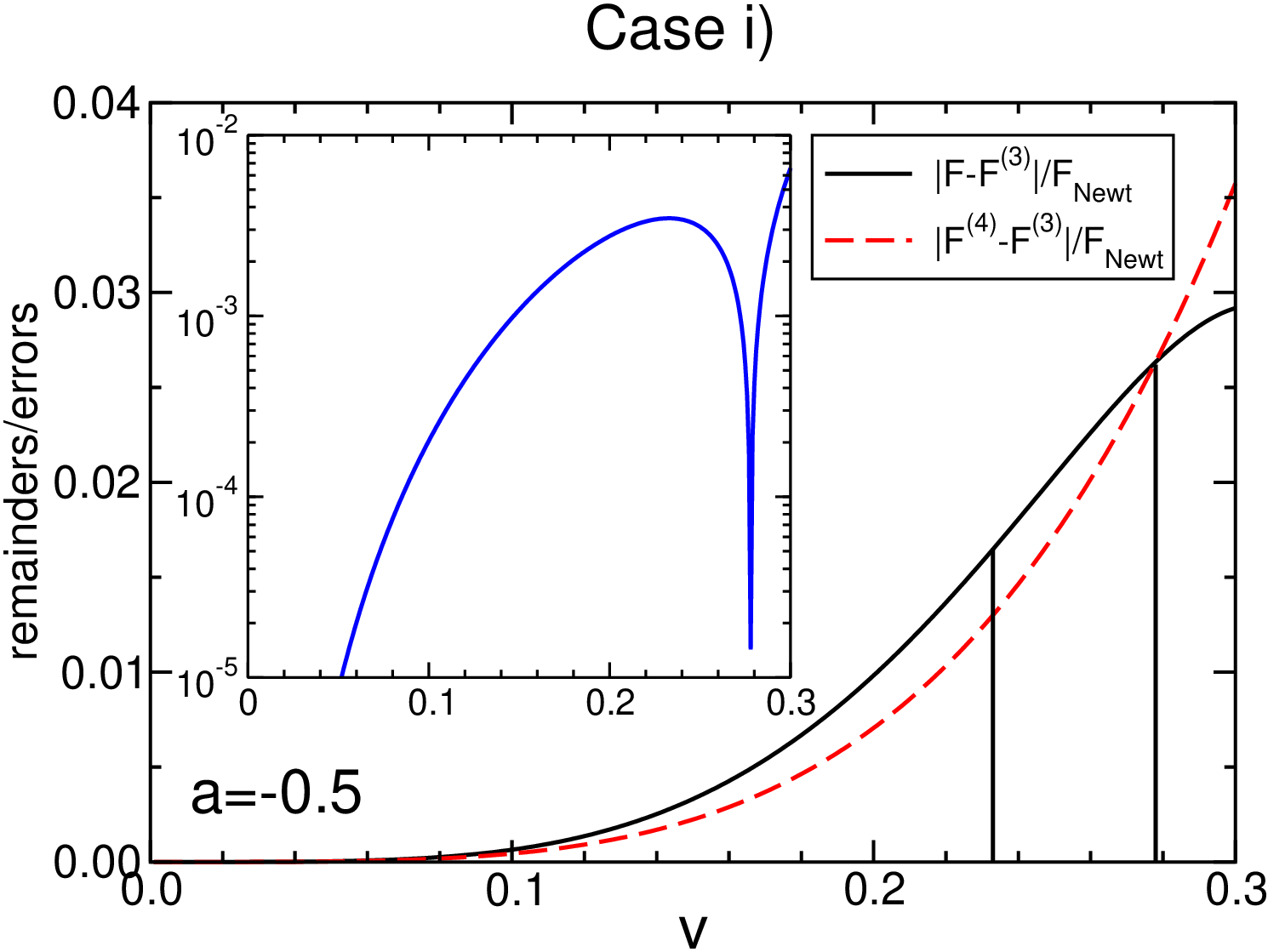}
\includegraphics[width=8.5cm,clip]{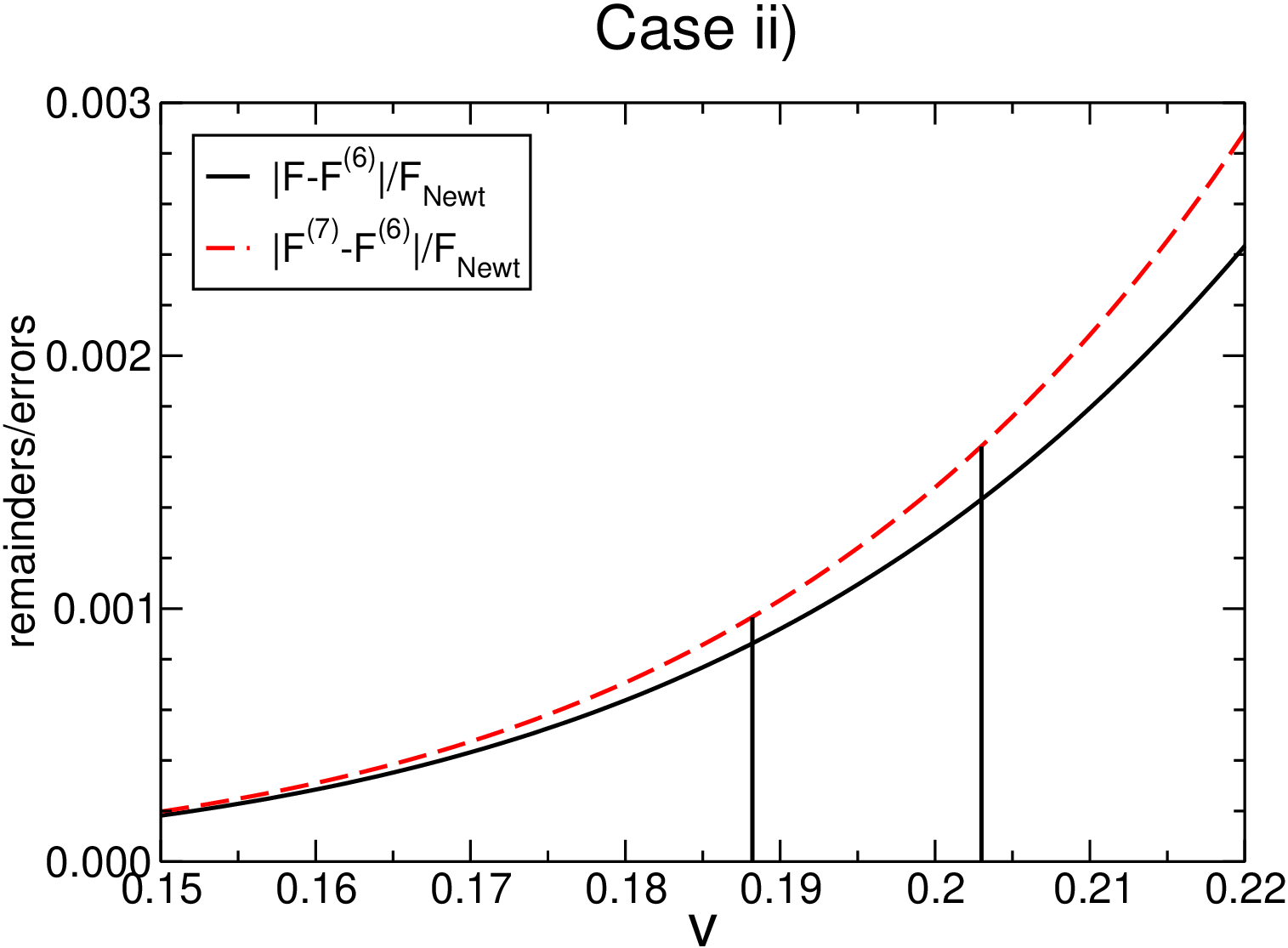}\\
\end{tabular}
\caption{\label{case} Left: absolute value of the remainder of the $N=3$ PN
  flux, $|F-F^{(3)}|/F_{\rm Newt}$ (solid line), and the $N=4$ term
  $|F^{(4)}-F^{(3)}|/F_{\rm Newt}$ (dashed red line). The inset shows the
  modulus of their difference, Eq.~(\ref{deltaN}), in a semilogarithmic
  scale. Right: same as the left panel, but for the $N=6$ remainder and $N=7$
  term. All curves in this plot refer to the counterrotating case with spin
  $a=-0.5$. The lower ($v_l$, more conservative) and upper ($\bar{v}$, less
  conservative) edges of the region of validity are (somewhat conventionally)
  delimited by the vertical lines, as explained in the main text.}
\end{figure*}

This figure provides some clues about the edge of the region of validity of
the PN approximation. For corotating orbits (left panel of Fig.~\ref{result}),
as the spin increases from zero to $a=0.9$, the higher-order PN approximants
start to deviate from numerical results at lower values of $v/v_{\ISCO}$: this
happens roughly when $v/v_{\ISCO}\simeq 0.6$ for $a=0$, and when
$v/v_{\ISCO}\simeq 0.4$ for $a=0.9$. This leads us to naively expect a
shrinking of the region of validity of the PN approximation as a function of
positive $a$. This expectation will be validated (at least qualitatively) in
Section \ref{sec:validity}: cf. the bottom-right panel of Fig.~\ref{coNv}
below.

At first sight, the results for counterrotating orbits (right panel of
Fig.~\ref{result}) seem surprisingly good. In particular, the 3PN
approximation (dash-dot-dotted, orange line) is almost indistinguishable from
the numerical result all the way up to $v=v_{\ISCO}$ when the spin is large.
Such a good performance is simply because of the well-known,
monotonically-increasing behavior of $v_{\ISCO}$ with spin, with a minimum as
$a \to -1$ (cf. Fig.~\ref{visco}). Since counterrotating orbits probe a
smaller range in $v/c$ (up to $v/c\sim 0.35$ for fast-spinning BHs), the PN
approximation is more accurate. Unfortunately, prograde accretion is likely to
be more common than retrograde accretion in astrophysical settings (see
e.g.~\cite{Berti:2008af}). Moreover, the 3.5PN and 4PN approximants are
significantly {\it worse} than the 3PN one at $a=-0.99$. This is consistent
with the PN series being an asymptotic expansion, as one of the characteristic
features of the latter is that beyond a certain optimal order, higher-order
approximants become less accurate~\cite{Bender:1978}.

\section{\label{sec:validity}Region of validity}

Let us now turn to determining the region of validity of the PN approximation
for different values of the BH spin. For a complete review of
asymptotic approximation techniques we refer the reader to \cite{Bender:1978};
Paper I presents a short introduction to the topic in the present context. As
explained in those references, the edge of the region of validity is
determined by the approximate condition
\be
\label{E-order}
{\cal{O}}(F - F^{(N)}) =
{\cal{O}}(F^{(N+1)} -F^{(N)})\,,
\ee
where $F$ denotes the ``true'' (numerical) result for the GW energy flux and
$F^{(N)}$ denotes the $N$-th order PN approximation.

An inherent and {\emph{intrinsic}} ambiguity is contained in
Eq.~\eqref{E-order}, encoded in the order symbol. This makes {\emph{any}}
definition of the region of validity of an asymptotic series somewhat
imprecise.  As shown in Fig.~\ref{case} (or in Fig.~8 of Paper I), there are
two qualitatively different scenarios:
\begin{enumerate}
\item[{\bf{\emph{i)}}}] Left panel of Fig.~\ref{case}: The next-order term
  $|F^{(N+1)} -F^{(N)}|$ starts off smaller than the remainder $|F -
  F^{(N)}|$, but eventually they cross and separate. We can then
  estimate the edge of the region of validity $\bar{v}$ by solving 
  $\delta^{(N)}(\bar{v})=0$, where
\be\label{deltaN}
\delta^{(N)}(v)\equiv \left||F-F^{(N)}|-|F^{(N+1)}-F^{(N)}|\right|\,
\ee
is the {\emph{error function}}. If we also define a more conservative {\em
  lower edge} of the region of validity, $v_{l}$, as the point where
\be\label{delmax}
\left.\frac{d\delta^{(N)}(v)}{dv}\right|_{v_l}=0\,,
\ee
we can then introduce an {\emph{uncertainty width}} of the region of validity:
$\delta\bar{v}\equiv \bar{v}-v_l$; see the inset of the left panel of
Fig.~\ref{case}.
\item[{\bf{\emph{ii)}}}] Right panel of Fig.~\ref{case}: The remainder and the
  next-order term are of the same order for sufficiently low velocities, until
  eventually the curves separate for larger velocities. This situation is the
  rule, rather than the exception, for the problem we consider in this
  paper. When this happens, method i) cannot be applied, because
  $\delta^{(N)}(v)=0$ has no solutions. Given the approximate nature of the
  order relationship in Eq.~(\ref{E-order}), we can {\em define} the region of
  validity as the point $\bar{v}$ such that $\delta^{(N)}(\bar{v})=\delta_0$,
  where $\delta_0$ is some given tolerance defined below.
\end{enumerate}

\begin{figure*}[htb]
\begin{center}
\begin{tabular}{ccc}
\includegraphics[width=6cm,clip=true]{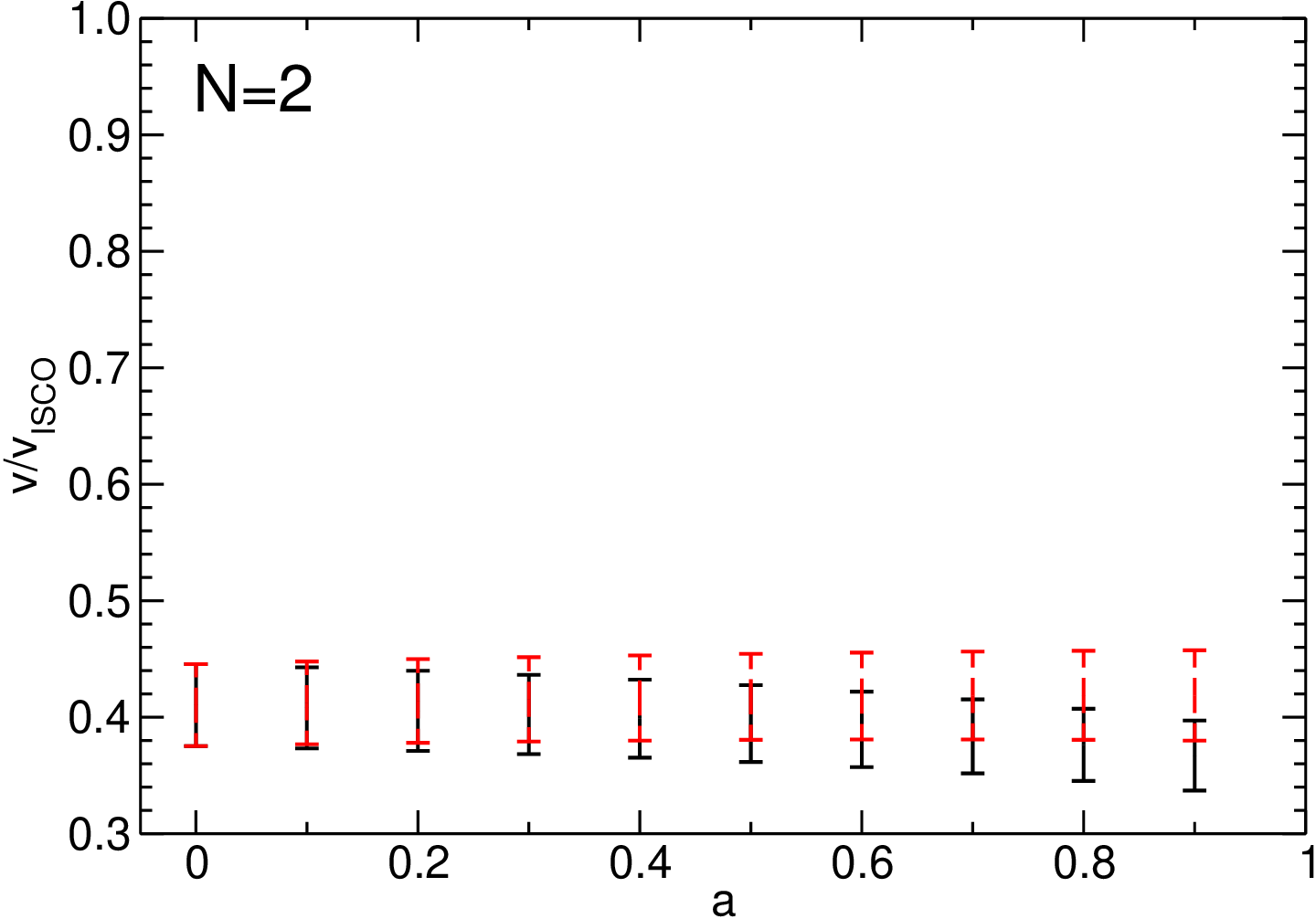}
&\includegraphics[width=6cm,clip=true]{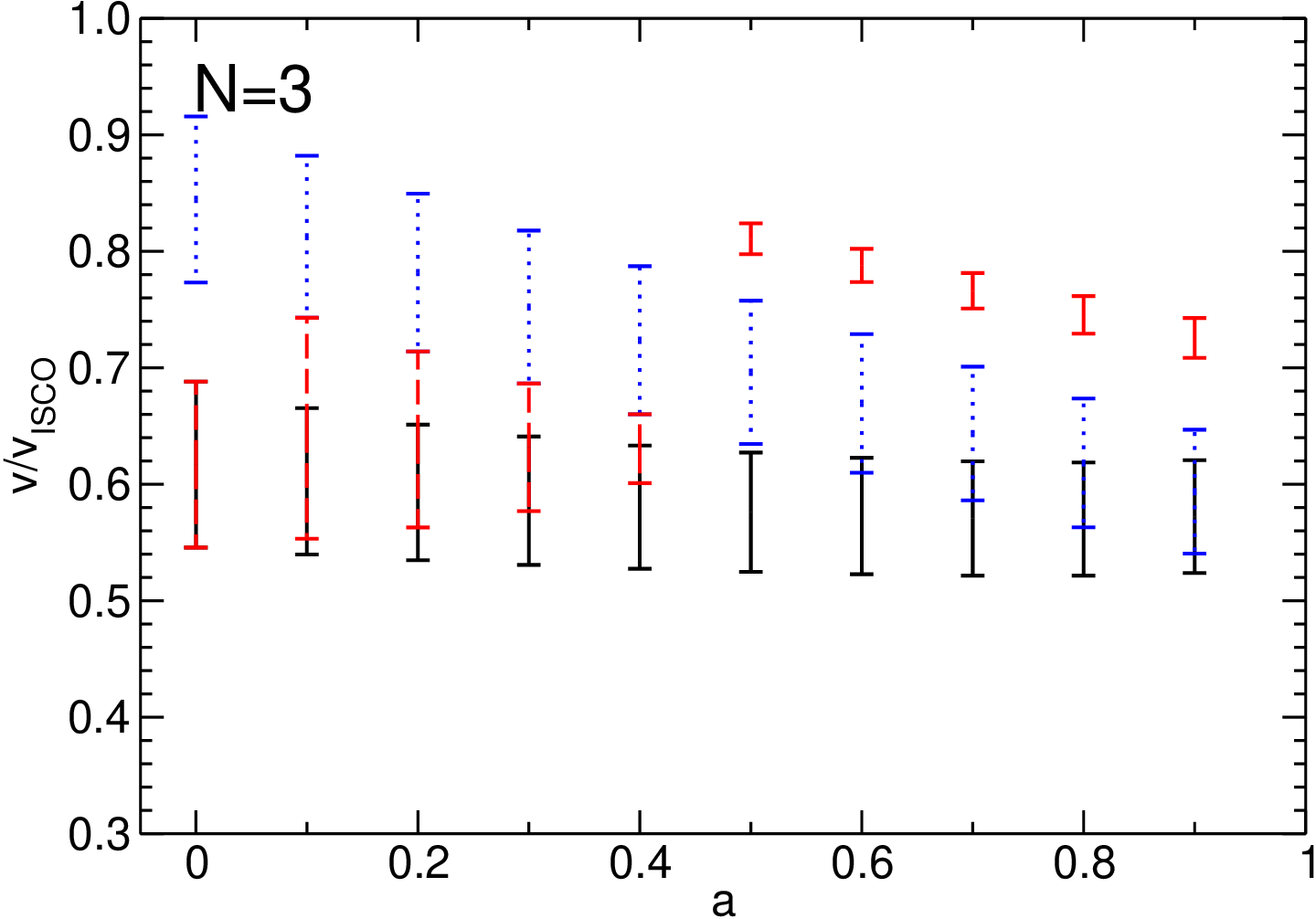}
&\includegraphics[width=6cm,clip=true]{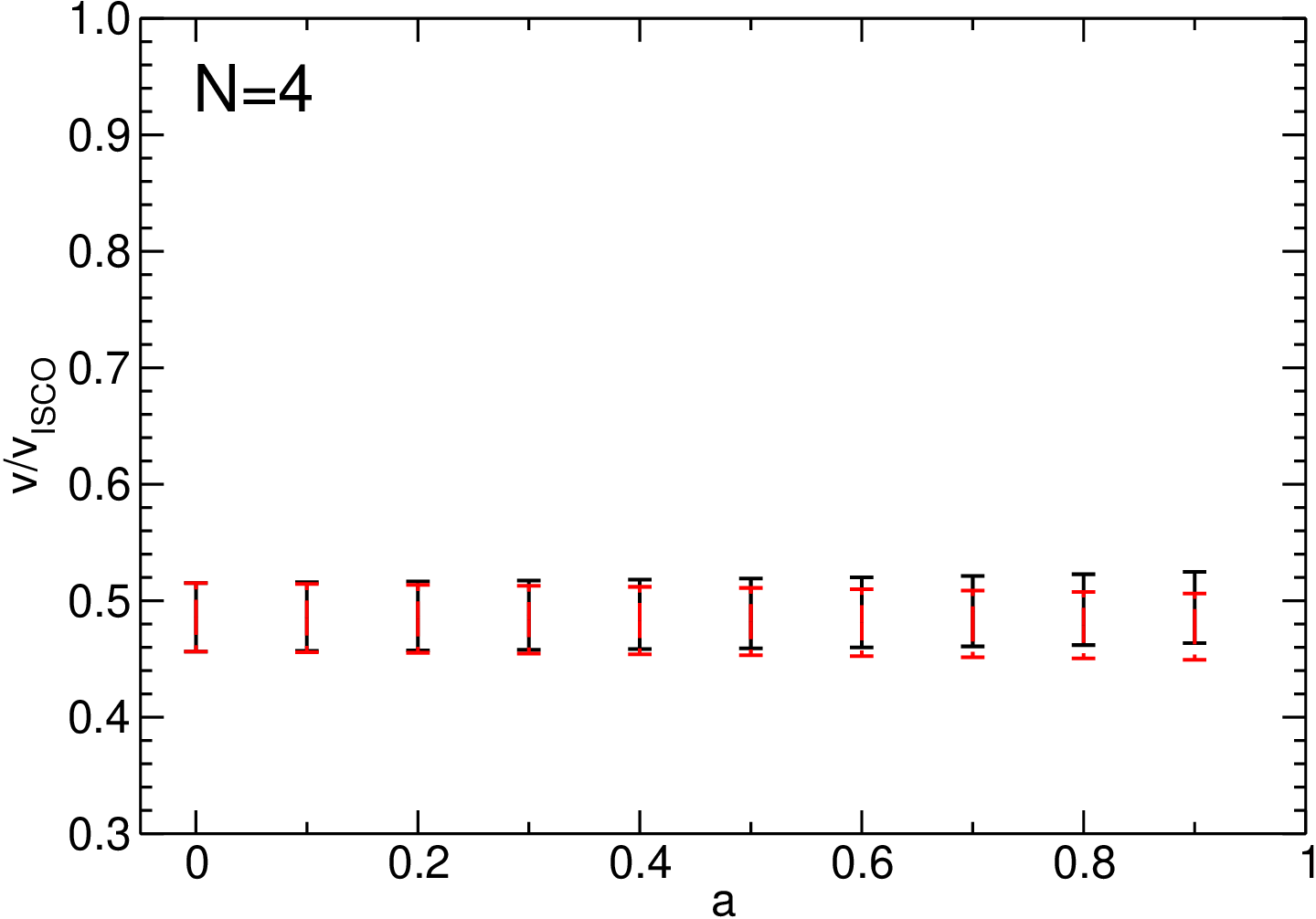}\\
\end{tabular}
\begin{tabular}{cc}
\includegraphics[width=6cm,clip=true]{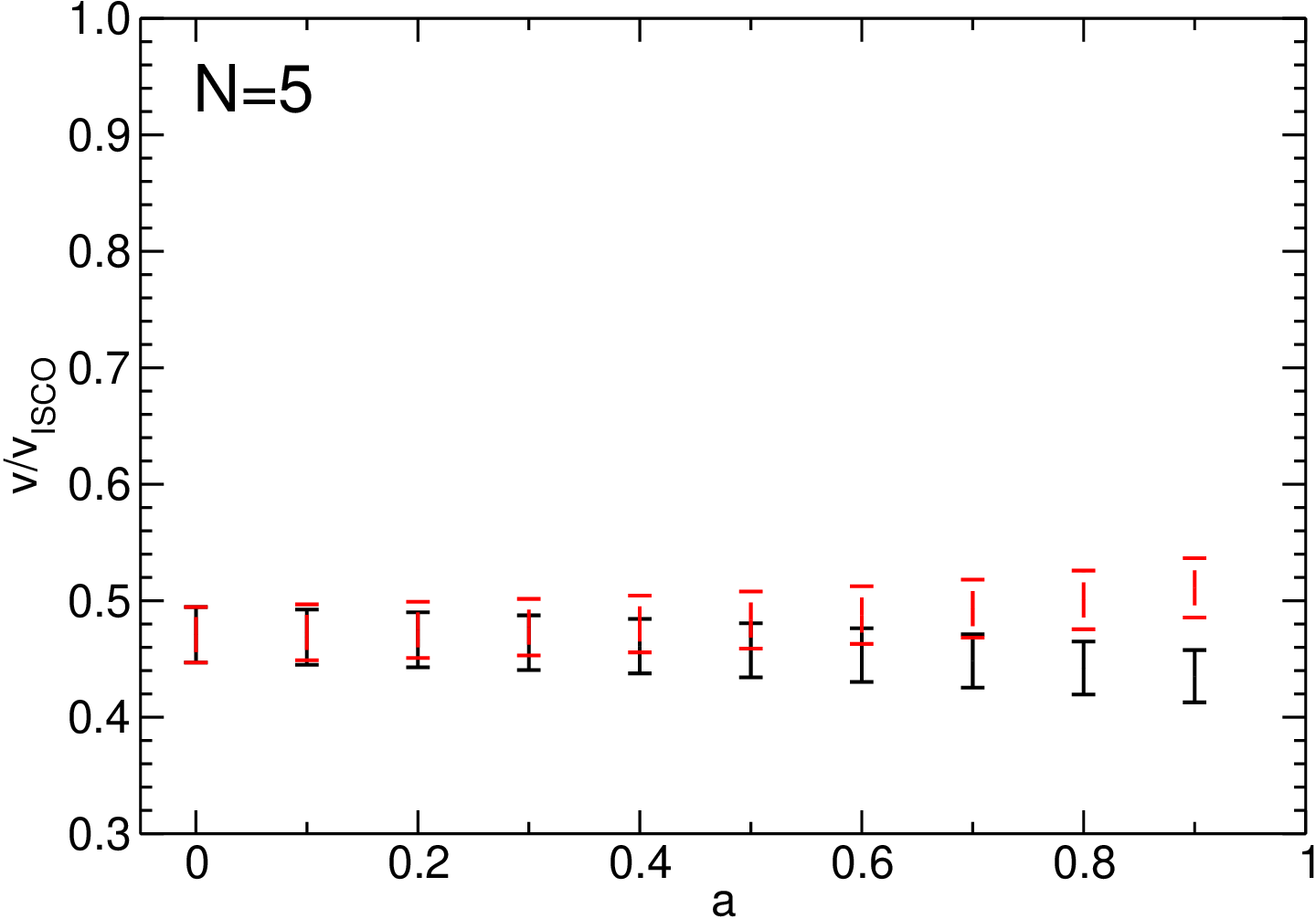}
&\includegraphics[width=6cm,clip=true]{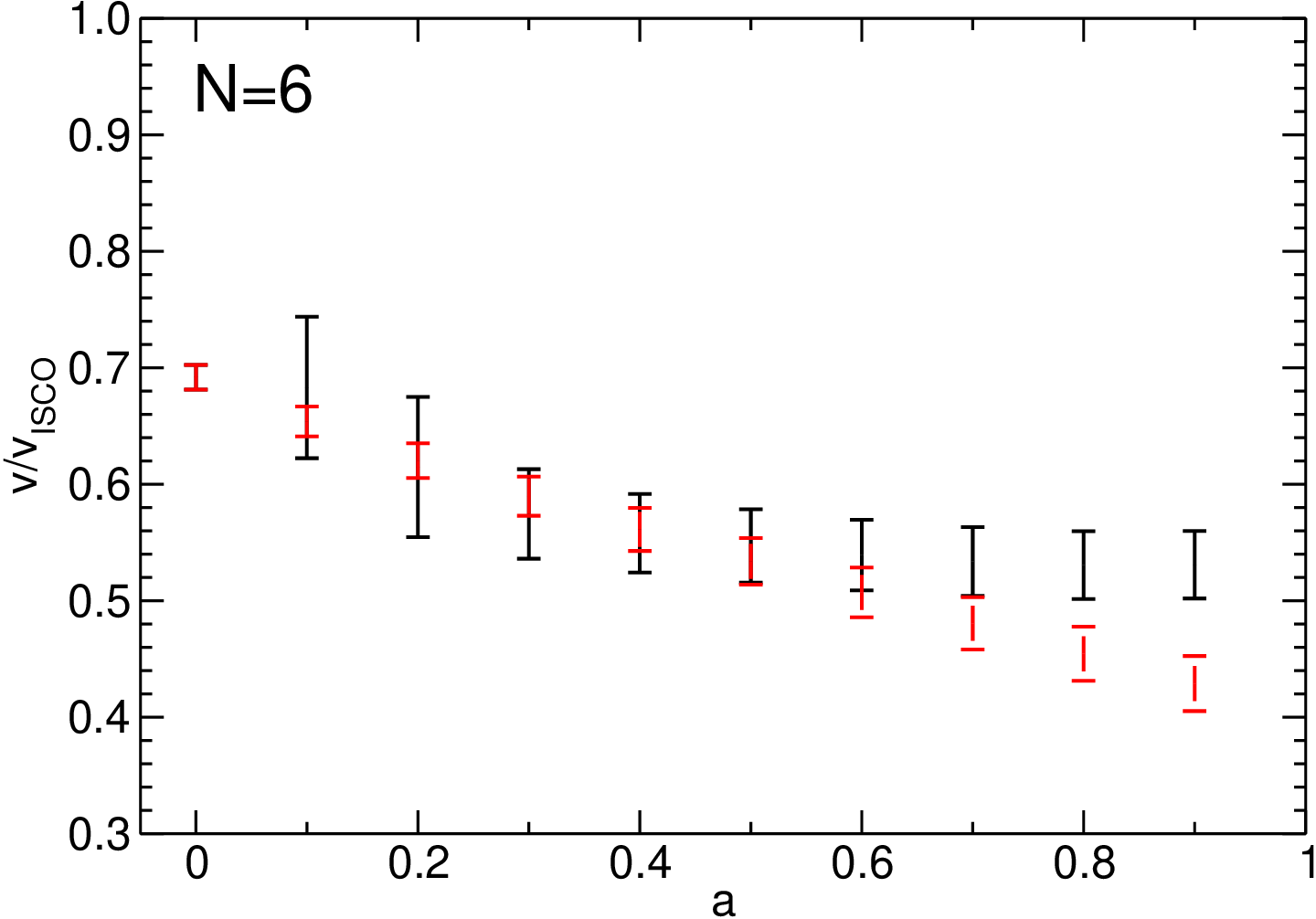}\\
\end{tabular}
\caption{\label{coNv} Edge of the region of validity as a function of $a$ for
  different PN orders, in the corotating (black, straight line) and
  counterrotating (red, dashed line) cases. The blue, dashed lines for $N=3$
  refer to the counterrotating case, and they were obtained by an alternative
  method (see the discussion around Fig.~\ref{fig:app} below).}
\end{center}
\end{figure*}

Higher-order approximations should be sensitive to a smaller tolerance, which
implies that $\delta_0$ cannot be set arbitrarily.  Instead, $\delta_0$ should
be given by the error in the difference between the $N$th remainder and the
$(N+1)$th-order term.  This error is presumably of the order of the error in
the $(N+1)$th-order term, and it can be estimated by the $(N+2)$th-order term.
The imprecision of the order symbol is now encoded in the fact that $\delta_0$
depends on $v$. We can try to estimate its value by evaluating the $(N+2)$th
order term in the middle of the allowed range, that is, at $v_{\ISCO}/2$:
\be\label{delta0}
\delta_0 = \left| F^{(N+2)}(v_{\ISCO}/2)-F^{(N+1)}(v_{\ISCO}/2) \right|\,.
\ee
This estimate of $\delta_0$ is not exact, so we can try to provide a more
conservative lower edge of the region of validity, $v_{l}$, by
imposing the condition\footnote{Note that in the Erratum of Paper I we impose
  a slightly different condition: $\delta(v_l)=\delta_0/2$, $\delta(\bar
  v)=2\delta_0$.} $\delta^{(N)}(v_l)=\delta_0/2$. We can then define an
uncertainty on the region of validity $\delta\bar{v}=|\bar{v}-v_l|$. This is
illustrated pictorially by the vertical lines in the right panel of
Fig.~\ref{case}.

Let us now discuss the behavior of the edge of the region of validity as a
function of the PN order $N$ and of the BH spin $a$.  The corotating
and counterrotating regions of validity and the associated errors are
shown in Fig.~\ref{coNv} with solid black (dashed red) error bars for
corotating (counterrotating) orbits, respectively.

Let us first consider the corotating case (solid black error bars). All
results were obtained using method ii) above. At any fixed PN order, the
normalized region of validity $v/v_{\ISCO}$ remains roughly constant as a
function of $a$. With a few exceptions, the most conservative estimate $v_l$
(lower edge of the error bars in the plots) is typically in the range
$v/v_{\ISCO}\in[0.3\,,0.6]$. This is consistent with the left panel of 
Fig.~\ref{result}, where we see that all PN approximations 
(including high-order ones) peel off from the numerical flux in this range.

These figures allow us to arrive at an interesting conclusion.  When we recall
that $v_{\ISCO}$ increases with $a$ (cf. Fig.~\ref{visco}), the figures
suggest that spin-dependent corrections in the PN expansion of
Eq.~(\ref{PNexp}) are effective at pushing the validity of the PN expansion to
higher values of $v/c$. However, there is an intrinsic limit to what is
achievable, which is determined instead by $v/v_{\ISCO}$, and roughly
independent of $a$. In the range $a\in[0.3\,,0.9]$, $v_{\ISCO}$ increases from
$\simeq 0.444$ to $\simeq 0.609$.  Therefore the region of validity for the
orbital velocity is approximately in the range $v/c \in [0.44\times
  0.3\,,0.61\times 0.6] \sim [0.13\,,0.37]$.

\begin{figure*}[htb]
\begin{center}
\begin{tabular}{ccc}
\includegraphics[width=6cm,clip=true]{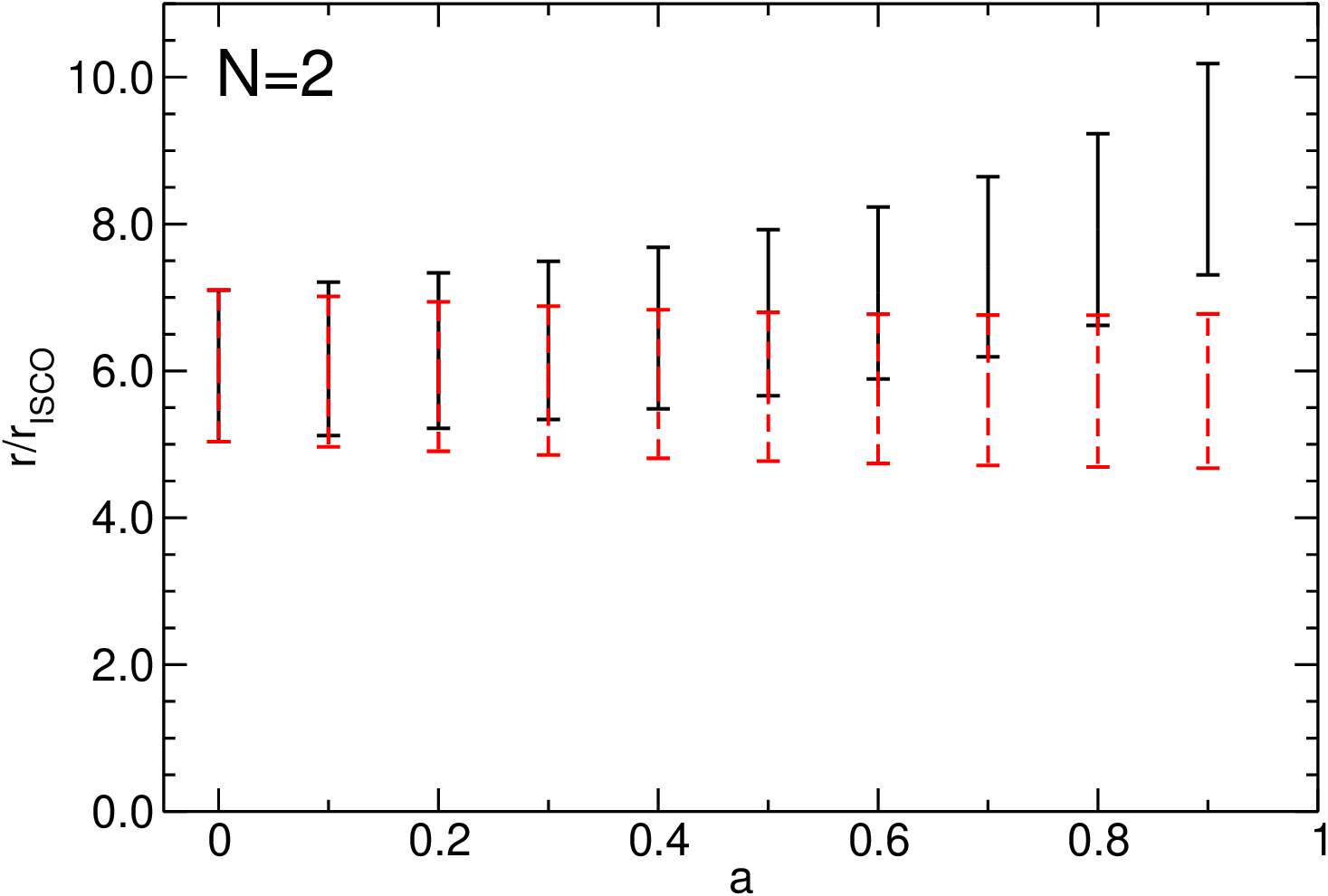}
&\includegraphics[width=6cm,clip=true]{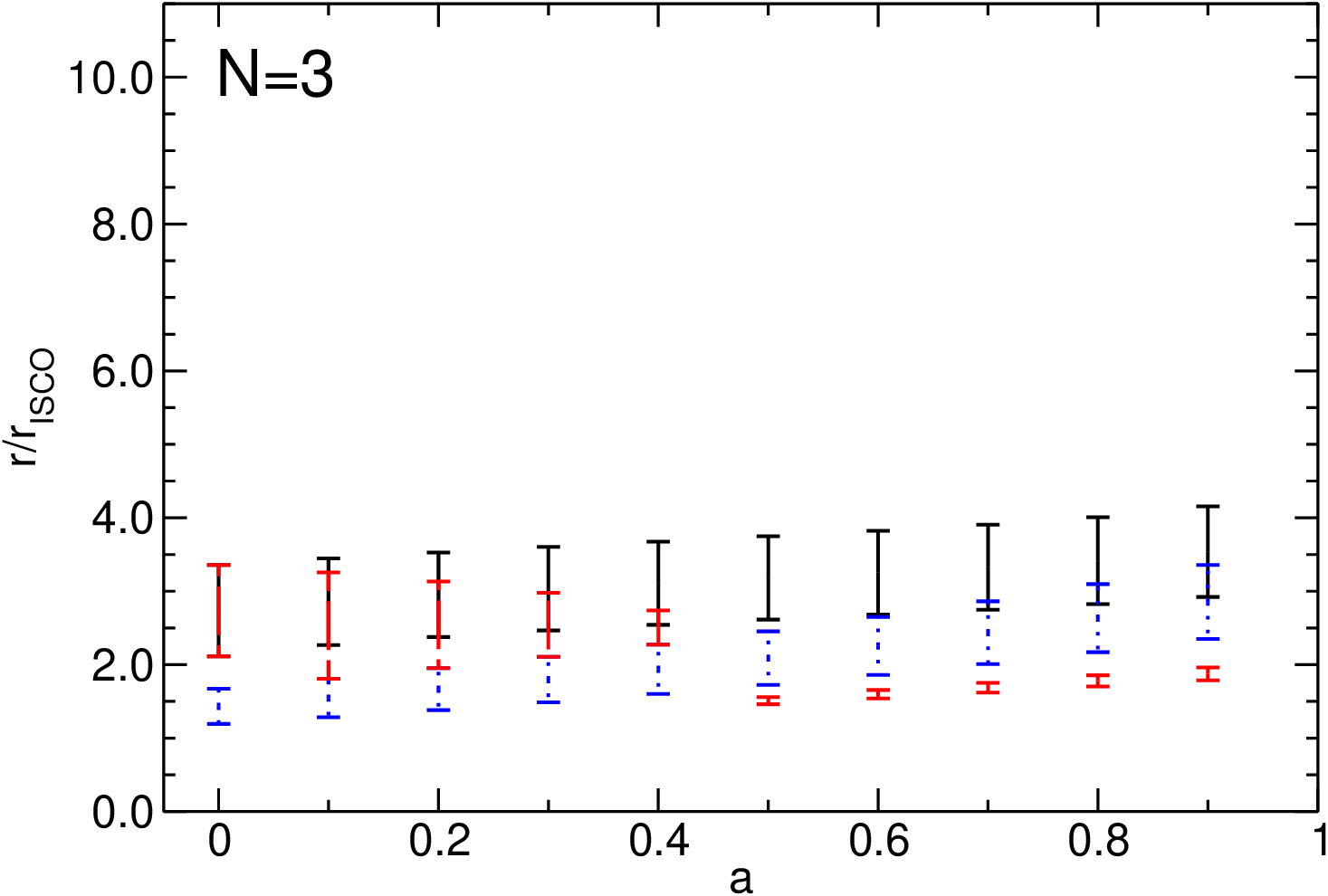}
&\includegraphics[width=6cm,clip=true]{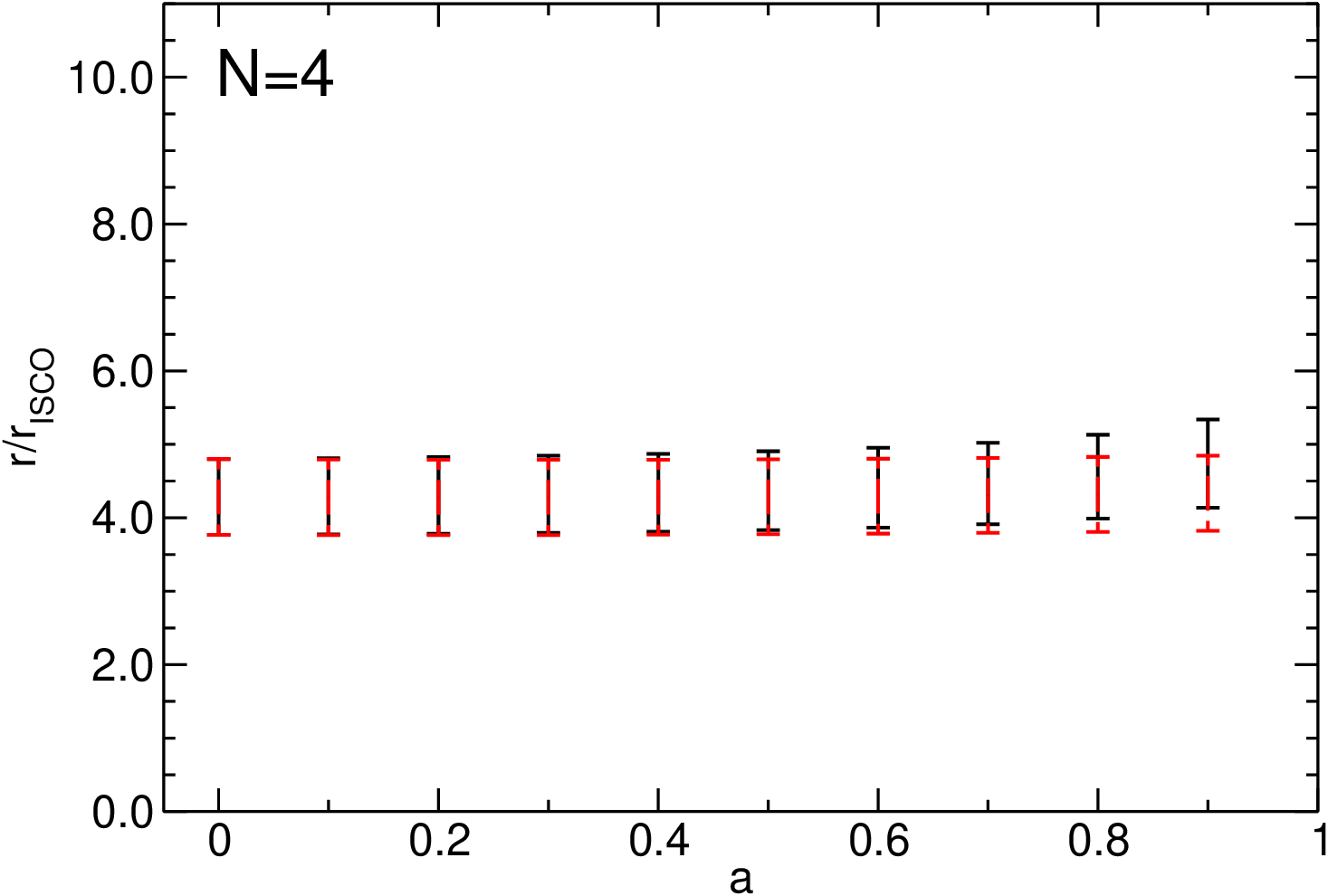}\\
\end{tabular}
\begin{tabular}{cc}
\includegraphics[width=6cm,clip=true]{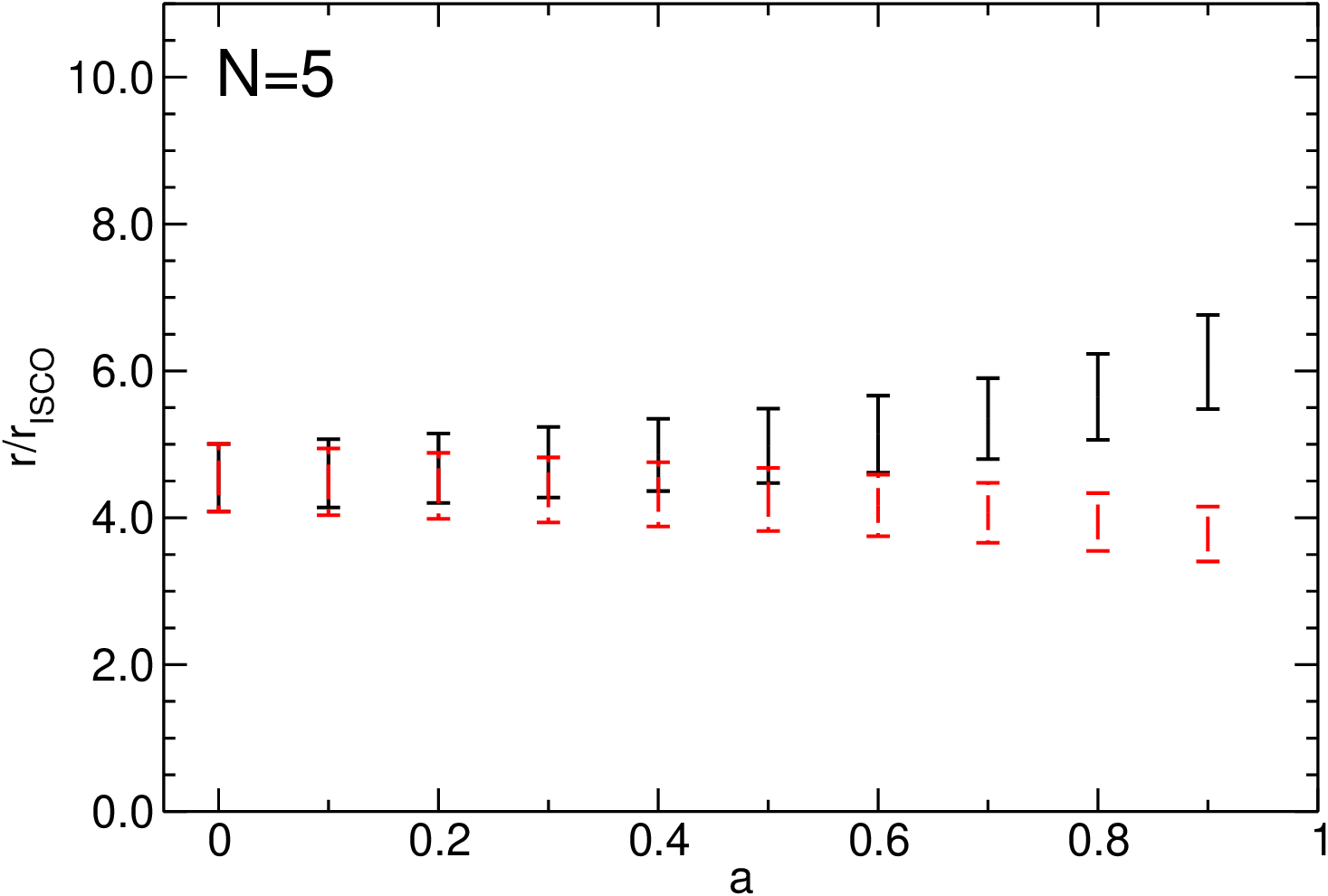}
&\includegraphics[width=6cm,clip=true]{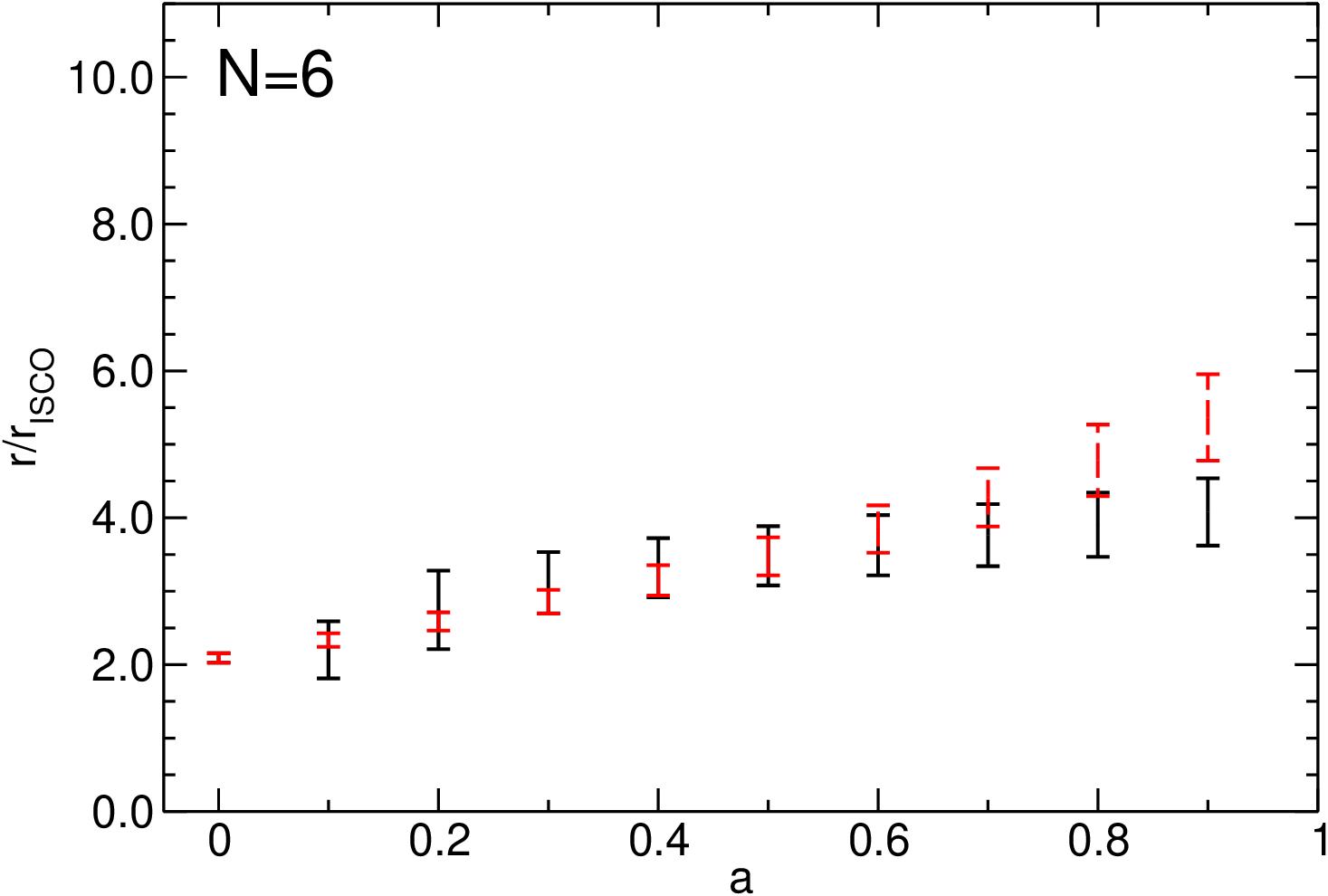}\\
\end{tabular}
\caption{\label{coNr} Edge of the region of validity expressed in terms of the
  Boyer-Lindquist radius for different PN orders, in the corotating (black,
  straight line) and counterrotating (red, dashed line) cases. The blue,
  dashed lines for $N=3$ refer to the counterrotating case, and they were
  obtained by an alternative method (see the discussion around
  Fig.~\ref{fig:app} below).}
\end{center}
\end{figure*}

Let us now focus on the counterrotating case, i.e. on the dashed red error
bars in Fig.~\ref{coNv}, which, again, were determined using method ii)
above. The only exception is the case $N=3$ (corresponding to the right panel
of Fig.~\ref{case}), that we will discuss separately below.  As in the
corotating case, the region of validity shrinks mildly or remains roughly
constant as $|a|$ increases. For $N=6$ the region of validity
shrinks faster with increasing spin.

The edge of the region of validity can also be presented in terms of the
Boyer-Lindquist radius of the particle's circular orbit. The corresponding
plots for the corotating and counterrotating cases are presented, for
completeness, in Fig.~\ref{coNr}. The ISCO radius $r_{\ISCO}$ is a
monotonically decreasing function of the spin (or of $v_{\ISCO}$), so, quite
naturally, the trend as a function of $a$ is the opposite of what we observed
for velocities in Fig.~\ref{coNv}. Our results consistently suggest that the
region of validity of the PN approximation cannot be extended all the way down
to the ISCO, contrary to a rather common assumption in GW data
analysis. Instead, one should use care when using, for example, the 2PN
approximation for $r/r_{\ISCO} < 4$, as in that regime higher-order PN
terms cannot be neglected (this is particularly true for rapidly rotating BHs
in prograde orbits). Our results suggest that a safer choice would be to
truncate all analyses at $r/r_{\ISCO} = 6$, which ranges between $r/M \in
[6\,,54]$ depending on the BH spin, unless one is dealing with approximants
more accurate than Taylor expansions.

\begin{figure}[htb]
\begin{center}
\includegraphics[width=8.5cm,clip=true]{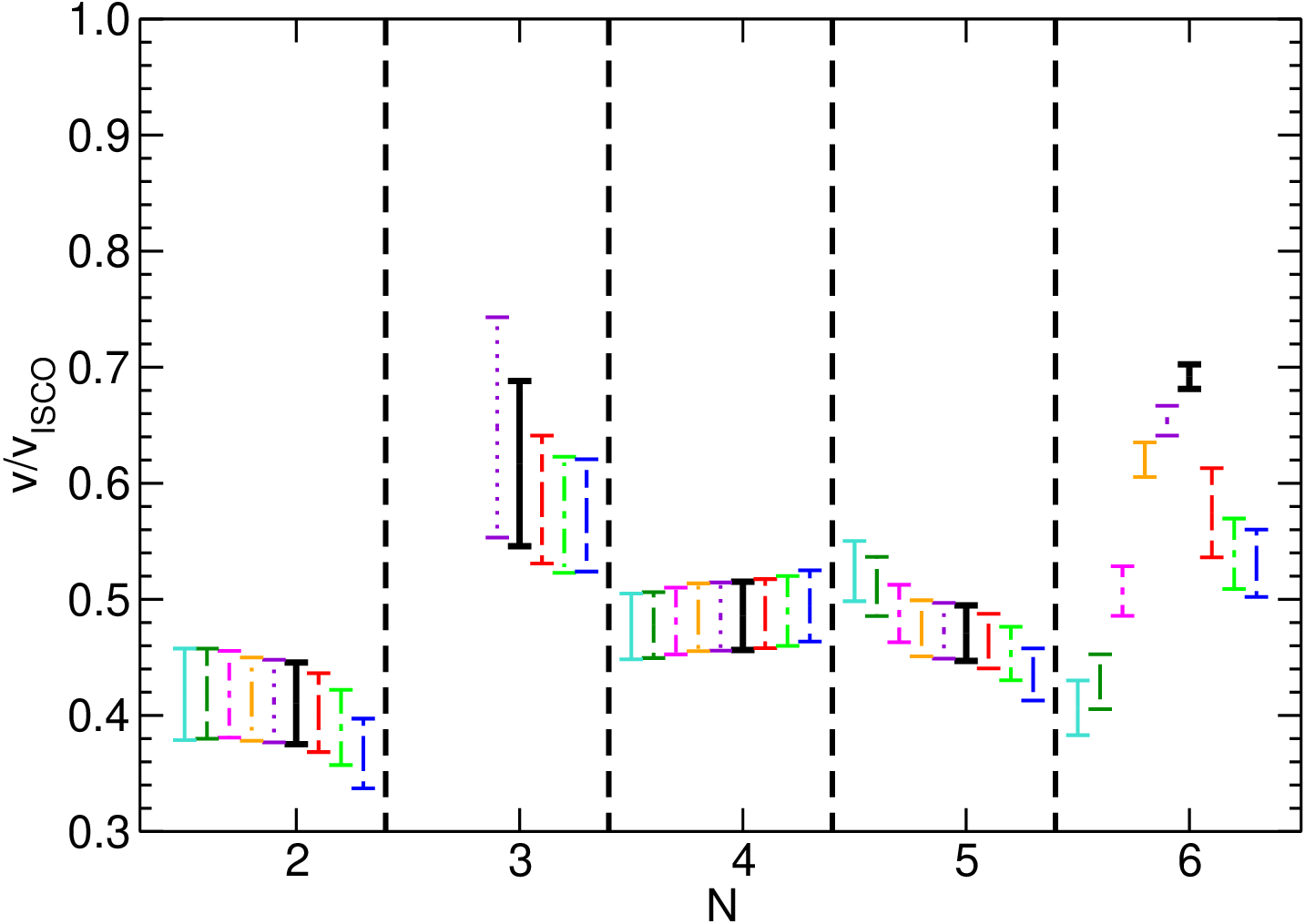}
\caption{\label{fixN} Edge of the region of validity as a function of $N$ for
  fixed values of $a$. The thick, black error bar corresponds to $a=0$,
  followed by $a=0.3\,,0.6\,,0.9$ to the right and
  $a=-0.1\,,-0.2\,,-0.6\,,-0.9\,,-0.99$ to the left.  Note that $N=3$ is a
  special case. For reasons discussed in the text, we do not give error bars
  for $N=3$ and $a<-0.1$.}
\end{center}
\end{figure}

Finally, one can also investigate how the edge of the region of validity
behaves with PN order. This is depicted in Fig.~\ref{fixN} for a set of fixed
values of $a$ (shown with different colors, as described in the caption). The
vertical dashed lines separate the different-$N$ orders. If we concentrate on
the nonspinning case (black), ignore the pathological case $N=3$ (discussed
below) and consider the conservative, lower end of the error bar, we see that
there is a maximum at $N=6$.  For larger values of $N$, $v/v_{\ISCO}$ would
consistently decrease, as found in Paper I.  In the spinning case, however,
this trend is not as clear, as at $N=6$ the edge of the region of validity is
rather sensitive to the spin value. Without higher-order terms in the PN
expansion, which would provide larger-$N$ points in this figure, one cannot
conclude whether $N=6$ is the optimal order of expansion in the spinning case.

Before moving on to the next Section, let us discuss the $N=3$ case for
counterrotating orbits in more detail. This is a special case, as noted by the
discontinuity in counterrotating orbits shown in Figs.~\ref{coNv}, \ref{coNr}.
The pathologies explained below are the reason why, in Fig.~\ref{fixN}, we only
plotted the counterrotating edge of the region of validity when the Kerr spin
parameter $|a|\leq 0.1$. Notice also that, when $N=3$, the error regions in
Fig.~\ref{fixN} are significantly larger than for any other $N$ value.  We
will discuss the reason for this below, but the impatient reader can skip to
the next section without loss of continuity.

\begin{figure}[htb]
\includegraphics[width=8.5cm,clip]{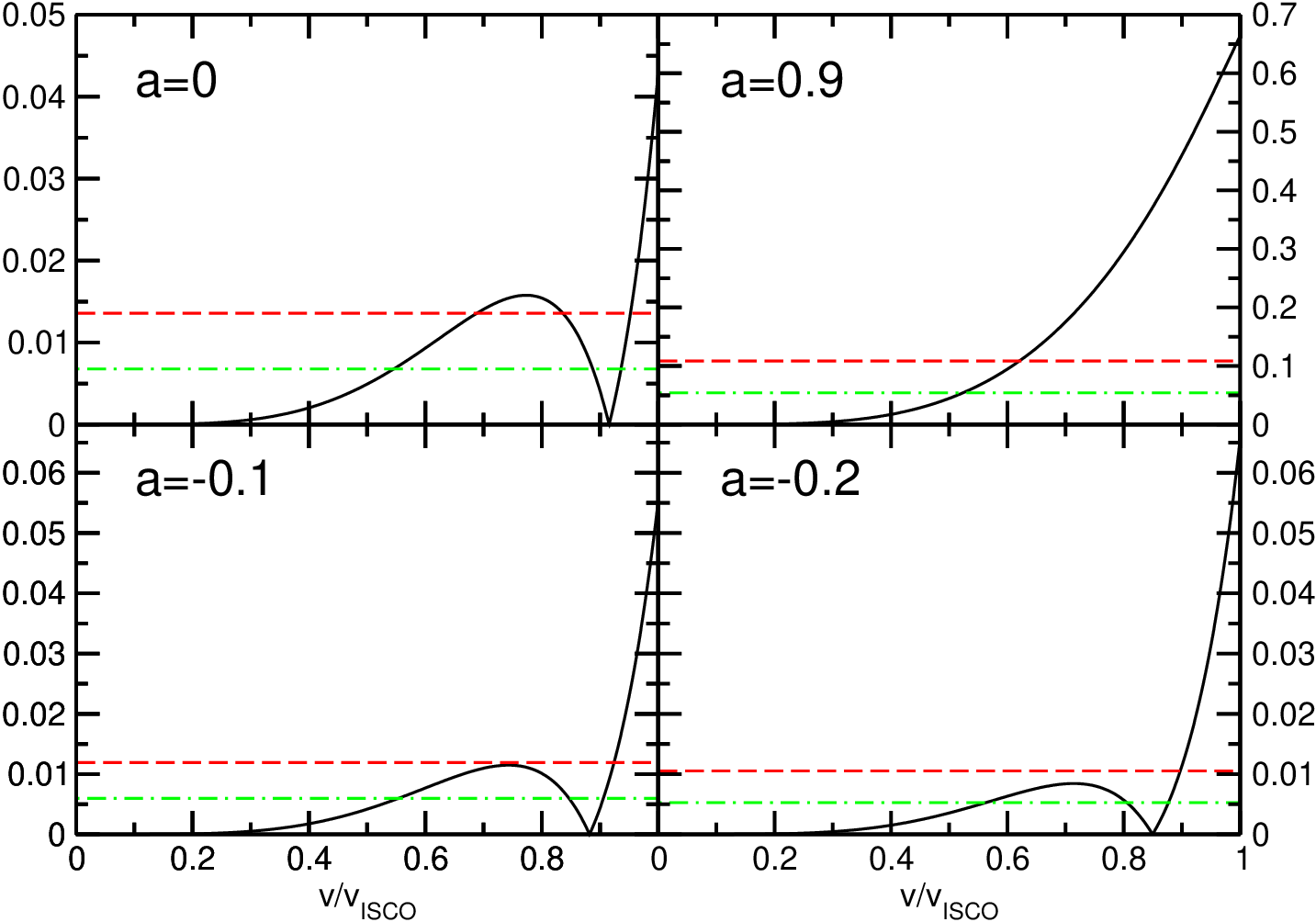}
\caption{\label{fig:app} The solid black line shows $\delta^{(N)}(v)$ for
  $N=3$. The horizontal dashed red (dot-dashed green) lines represent
  $\delta_0$, computed from Eq.~(\ref{delta0}), and $\delta_0/2$. See text for
  discussion.}
\end{figure}

Figure \ref{fig:app} clarifies the origin of the problem. When we use method
ii), the top margin of the edge of the region of validity is estimated as the
(smallest) value of $\bar{v}/v_{\ISCO}$ for which
$\delta^{(3)}(\bar{v})=\delta_0$. This condition corresponds to the leftmost
intersection of the horizontal dashed red line with the solid black line in
the plot.  Similarly, we determine the most conservative estimate of the edge
of the region of validity by considering the smallest $v_l$ such that
$\delta^{(3)}(v_l)=\delta_0/2$. This corresponds to the leftmost intersection
of the horizontal, dot-dashed green line with the solid black line. For
corotating orbits, as it happens, these intersections always exist. In fact,
the local maximum in $\delta^{(3)}(v)$ (which is located at $v/v_{\ISCO}\sim
0.8$ for $a=0$) moves to the {\it right} and becomes significantly larger as
$a\to 1$. For counterrotating orbits the trend is the opposite: the local
maximum moves to the left and decreases in magnitude. For a critical value of
the spin $a\simeq -0.1$, the red dashed line and the solid black line do not
intersect anymore. This is why in Figs.~\ref{coNv} and \ref{coNr} we only plot
the red-dashed (counterrotating) edge of the region of validity when $|a|\leq
0.1$. Of course, we can insist to identify $\bar{v}$ and $v_l$ as the smallest
values of $v$ such that $\delta^{(3)}(\bar{v})=\delta_0$,
$\delta^{(3)}(v_l)=\delta_0/2$. This procedure leads to the red, dashed error
bars in the $N=3$ panel of Figs.~\ref{coNv} and \ref{coNr}. Note that these
error bars are unnaturally small for $|a|>0.4$.

Another possible solution would be to switch to method i) when method ii)
fails. Now the upper margin of the edge of the region validity would be given
by the first zero of $\delta^{(3)}(v)$, and the lower margin would be
estimated by the condition given in Eq.~(\ref{delmax}). This results in the
blue, dotted error bars shown in the central top panel of Figs.~\ref{coNv} and
\ref{coNr}. These error bars are significantly more optimistic than the ones
we presented in the rest of the paper, but (in our opinion) their significance
is not as clear and well-justified as the rest of our results.

The problem discussed in this section concerns counterrotating orbits and
$N=3$. This is an exceptional case, and it does not affect the conclusions
drawn earlier in the paper. However we should remark, for completeness, that
similar pathologies occur for corotating orbits with $N=6$ when $0\leq a\leq
0.2$, and they may also occur at higher PN orders.

\begin{figure*}[htb]
\includegraphics[width=8.5cm,clip]{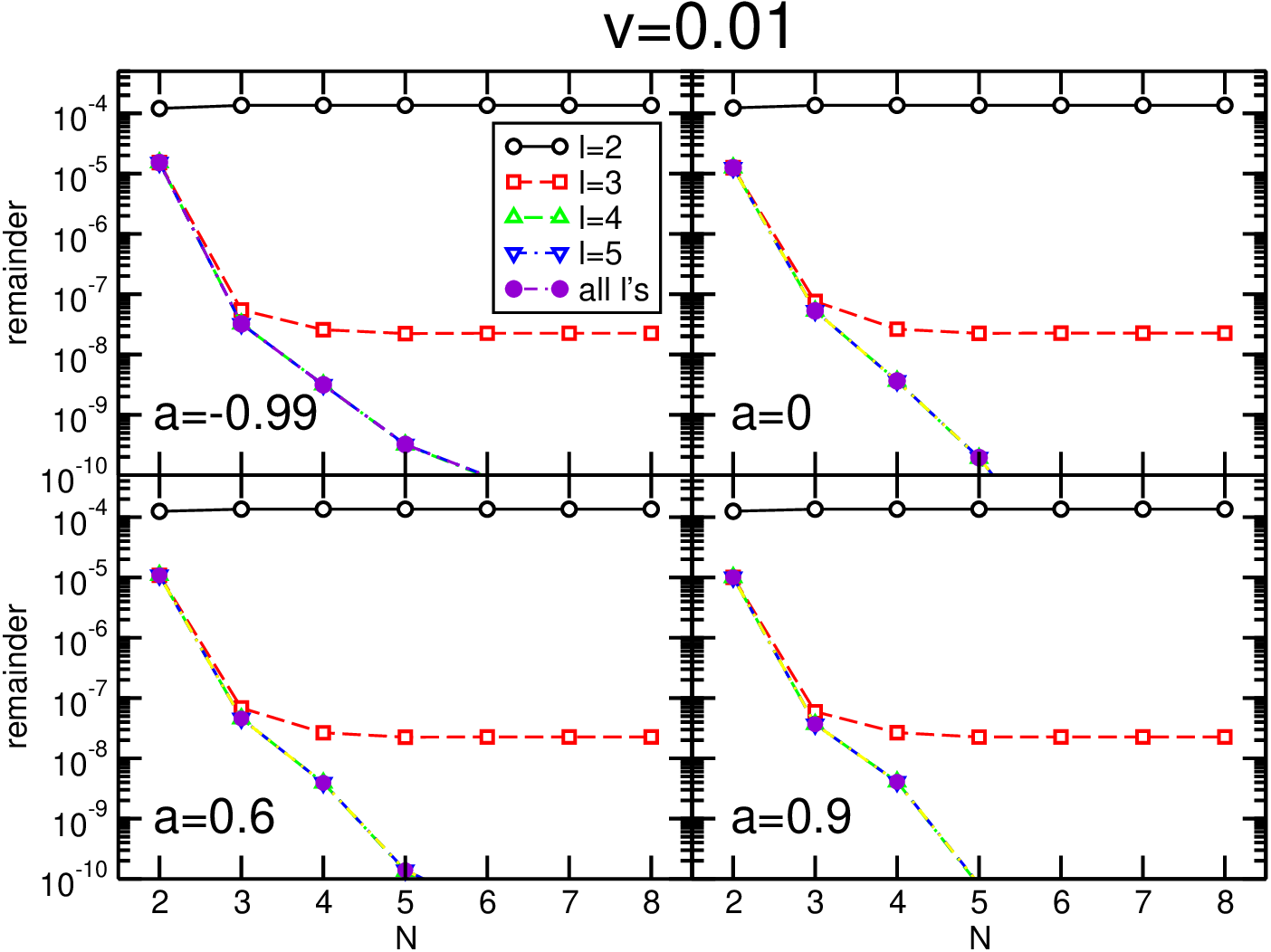}
\begin{tabular}{cc}
\includegraphics[width=8.5cm,clip]{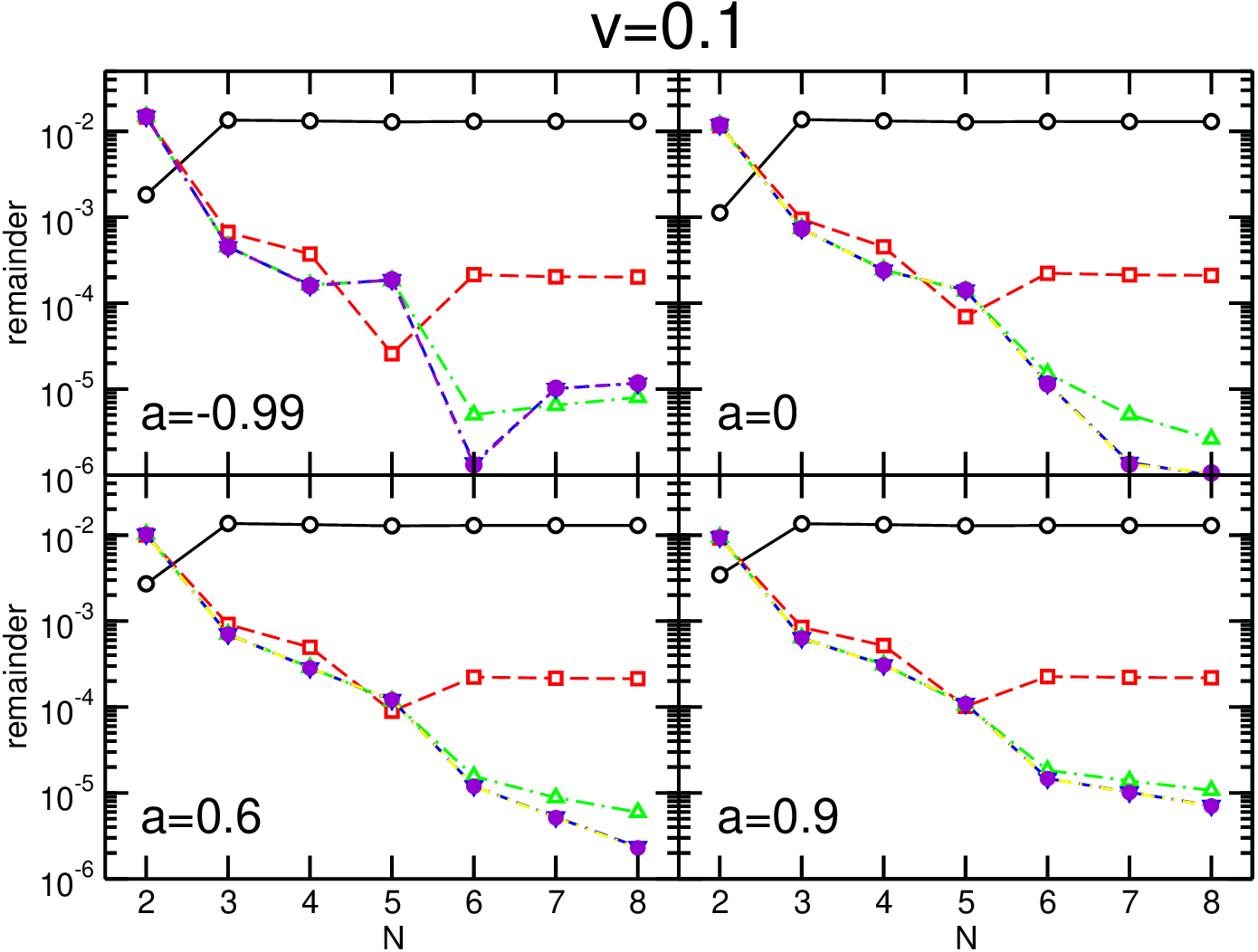}&
\includegraphics[width=8.5cm,clip]{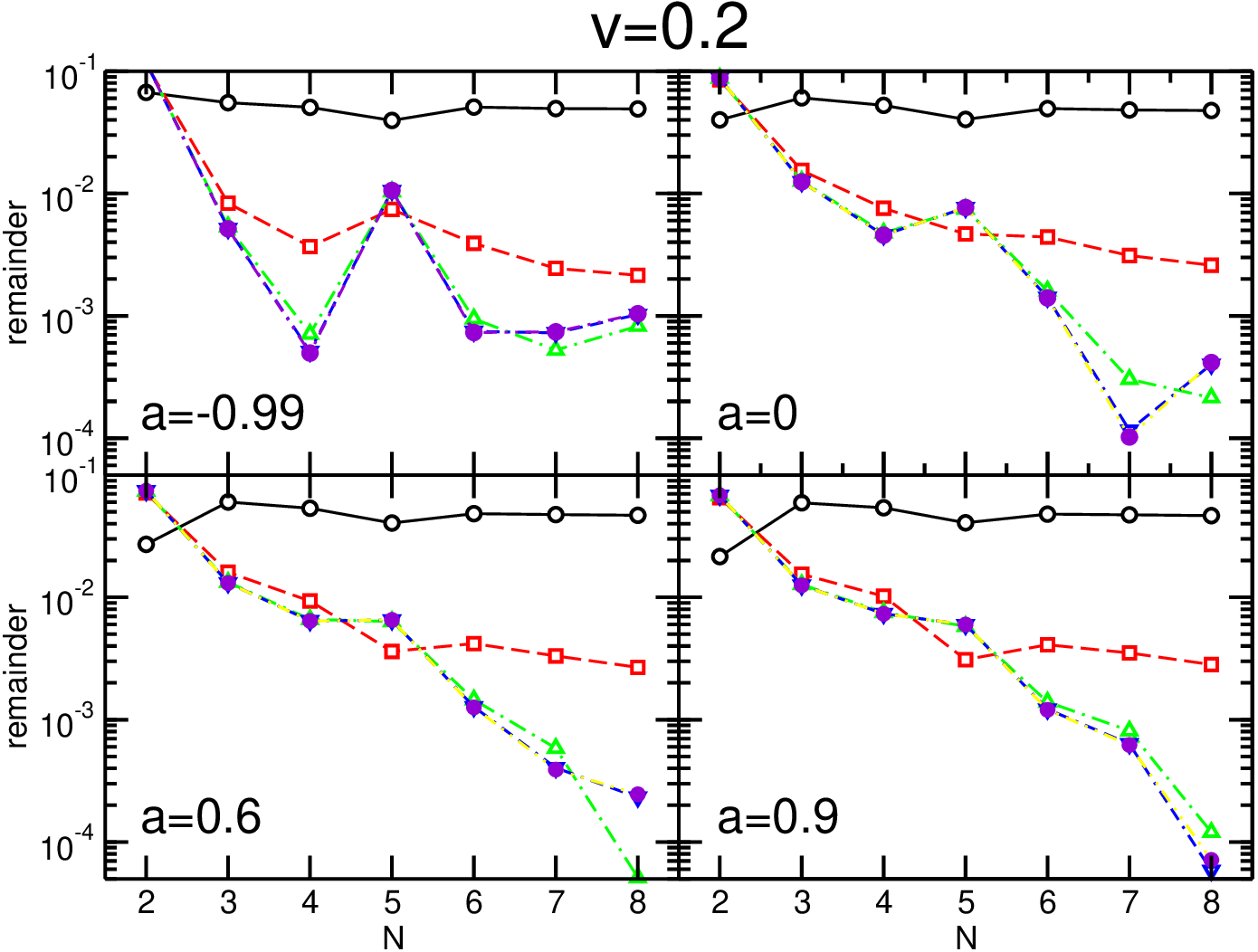}\\
\end{tabular}
\caption{\label{multipoles} Relevance of multipolar components up to $\ell=2$,
  $\ell=3$, $\ell=4$, $\ell=5$, and summing as many $\ell$'s as necessary for
  the relative accuracy of the Teukolsky code to be ${\cal O}(10^{-10})$ at
  any given velocity.}
\end{figure*}

\section{\label{sec:multipoles}Relevance of multipolar components as a
  function of spin}

Until now, we compared the PN approximation to numerical results that were
considered to be virtually ``exact''. This was justified because the Teukolsky
code computes as many multipoles in the angular decomposition of the radiation
as needed to achieve an accuracy of ${\cal{O}}(10^{-10})$ at any given orbital
velocity. While this is manageable in frequency-domain calculations, sometimes
accurate calculations of a large number of multipoles are not possible in
extreme mass-ratio {\it time-domain} codes, or in numerical relativity
simulations of comparable-mass binaries: cf. \cite{Berti:2007fi,Berti:2007nw}
for an analysis of multipolar decompositions of the radiation from
comparable-mass binaries and \cite{Pollney:2009yz,Pollney:2010hs} for more
recent numerical work to overcome these difficulties. As advocated in several
papers
\cite{Yunes:2008tw,Damour:2008gu,Bernuzzi:2010xj,2009GWN.....2....3Y,Yunes:2009ef,Yunes:2010zj,Pan:2010hz},
EMRIs provide a simple playground to study the number of multipolar components
required to reach a given accuracy in the PN approximation (or in one of its
resummed variants).

Figure~\ref{multipoles} shows a comparison of the convergence of
the multipolar decomposition versus the convergence of the PN expansion of the
energy flux. This plot generalizes Fig.~7 of Paper I in two ways: (i) it uses
more accurate numerical data, and (ii) it considers the effect of the central
BH spin on the number of multipolar components required to achieve a
given accuracy.

We fix three values of the orbital velocity ($v=0.01$, $v=0.1$ and $v=0.2$)
and we plot $F^{(\ell)}-F^{(N)}$, where $F^{(N)}$ is the $N$th approximant of
the PN energy flux and $F^{(\ell)}$ is the numerical energy flux truncated at
the $\ell$th angular multipole. Some features are immediately visible from
this plot:

\begin{itemize}

\item[(i)] Even at low orbital velocities ($v=0.01$), it is necessary to
  include multipolar components up to and including $\ell=4$ to achieve an
  accuracy better than $10^{-7}$ in the flux; on the other hand,
  including up to $\ell=5$ we obtain results that are as accurate as those
  that would be obtained including more multipoles.

\item[(ii)] For an orbital velocity $v=0.1$ ($v=0.2$) the best-known PN flux
  and numerical calculations always disagree at levels of $\sim 10^{-6}$
  ($\sim 10^{-4}$) or larger. This is obviously due to the slower,
  nonmonotonic convergence of the PN approximation in this regime. Some
  nontrivial features of the PN approximation are again well-visible here: for
  example, as pointed out repeatedly in this paper, when $a=-0.99$ and $v=0.1$
  the 3PN ($N=6$) expansion performs much better than higher-order expansions.
  This may well be accidental, and in fact it does not hold when $v=0.2$, as
  then $N=4$ is (most likely accidentally) better.

\item[(iii)] As a rule of thumb, the inclusion of multipoles up to and
  including $\ell=5$ seems necessary to achieve the level of accuracy of the
  best-known ($N=8$) PN expansion of the flux in the Kerr case. This
  conclusion is independent of the spin parameter $a$. In fact, $a$ has hardly
  any effect on the number of multipolar components that must be included in
  the flux to achieve a desired accuracy.
\end{itemize}

\section{\label{sec:concl}Conclusions}

We extended the method proposed in Paper I to determine the formal region of
validity of the PN approximation for quasicircular EMRIs of compact objects in
the equatorial plane of a Kerr BH. The boundary of the formal region
of validity is defined as the orbital velocity where the ``true'' error in the
approximation (relative to high-accuracy numerical calculations) becomes
comparable to the series truncation error (due to neglecting higher-order
terms in the series).

For quasicircular, equatorial Kerr EMRIs, the PN expansion is known up to 4PN,
and our estimate of the region of validity can only be pushed up to 3PN.  Our
main results are shown in Fig.~\ref{coNv} (in terms of orbital velocity) and
in Fig.~\ref{coNr} (in terms of orbital radius). At fixed but arbitrary spin
parameter $a$, the 3PN approximation has no obvious advantage when compared
with other PN orders. At fixed PN order $N$, Fig.~\ref{coNv} shows an interesting trend: 
when normalized by the ISCO velocity $v_{\ISCO}$, to a very good approximation
the region of validity does not depend on $a$.

We should emphasize that our results say nothing about the absolute accuracy
of the PN approximation: they only suggest relational statements between the
$N$th and the $(N + 1)$th-order approximations.  For velocities within the
region of validity of the asymptotic series, all we can say is that the $N$th
order approximation has errors that are of expected relative size. For larger
velocities, the $(N+1)$th- and higher-order terms become important, and should
not be formally neglected. If we can tolerate errors larger than those
estimated by the $(N + 1)$th order term (at the risk that higher-order
approximations may be less accurate than lower-order ones) we can surely use the
PN expansion beyond the realm of its formal region of validity. This, however, would
force us to lose analytic control of the magnitude of the error, as given by
the next order term.  The meaning of this caveat is well illustrated by the
counterrotating case with $a=-0.99$: it is clear from Fig.~\ref{result} that
the 3.5PN and 4PN approximations do {\it not} represent an improvement over
the 3PN approximation (and in fact perform quite badly) beyond the realm of
the region of validity.

Future work could concentrate on studying whether resummation techniques, such
as Pad\'e~\cite{Damour:1997ub,Damour:2000zb,Mroue:2008fu} or
Chebyshev~\cite{Porter:2007vk}, enlarge the formal region of validity, using
the methods developed here. This would allow us to identify optimal
resummation methods, which, in turn, affects resummed waveform models for
EMRIs, like the effective-one-body
approach~\cite{2009GWN.....2....3Y,Yunes:2009ef,Yunes:2010zj}.  Moreover, one
could attempt to establish whether there is a correlation between the accuracy
of the energy flux, as measured by the asymptotic methods developed here, and
the accuracy of the waveform model, as required by GW detectors.

\noindent
{\bf \em Note added in proof.} After submission of this paper, we learned that
Fujita has extended PN calculations of the energy flux up to 14PN for
nonrotating (Schwarzschild) BHs \cite{Fujita:2011zk}.

\acknowledgments 
We are grateful to Scott Hughes for comments on the manuscript and for
providing the numerical Teukolsky data for the energy flux, without which this
paper would not have been possible.  Z.Z. thanks Yanbei Chen for hospitality
at Caltech during December 2010 and January 2011. E.B. and Z.Z.'s research was
supported by the NSF under Grant No. PHY-0900735. N.Y. acknowledges support
from NASA through the Einstein Postdoctoral Fellowship PF9-00063 and
PF0-110080 issued by the Chandra X-ray Observatory, which is operated by the
SAO for and on behalf of NASA under contract NAS8-03060.

\end{document}